\documentclass[]{aa}
\usepackage{txfonts,graphicx,amssymb,subfigure}
\usepackage[normalem]{ulem}

\begin{document}
\title{Enhancement of magnetic fields arising from galactic encounters
}

   \author{D.Moss\inst{1}
          \and
          D. Sokoloff\inst{2}
          \and
          R. Beck\inst{3}
          \and
          M.Krause\inst{3}
          }

   \offprints{D.Moss}

   \institute{ School of Mathematics, University of Manchester, Oxford Road, Manchester, M13 9PL, UK
   \and
    Department of Physics, Moscow University, 119992 Moscow, Russia
               \and
    MPI f\"ur Radioastronomie, Auf dem H\"ugel 69, 53121 Bonn, Germany}

   \date{Received ..... ; accepted .....}

\abstract{Galactic encounters are  usually marked by
a substantial increase in synchrotron emission of the interacting galaxies when compared with the typical
emission from similar non-interacting galaxies. This increase is believed to be associated with an increase in
the star formation rate and the turbulent magnetic fields resulting from the encounter,
 while the regular magnetic field
is usually believed to decrease as a result of the encounter.}
{We attempt to verify these expectations. }
{We consider a simple, however rather realistic, mean-field galactic dynamo model where the effects of small-scale
generation are represented by random injections of magnetic field resulting from star forming regions. 
We represent an encounter by the introduction of large-scale streaming velocities
and by an increase in small-scale magnetic field injections. The latter 
describes the effect of an increase in the star formation rate caused by
the encounter. }
{We demonstrate that large-scale streaming, with associated deviations in the rotation 
curve, can result in an 
enhancement of the anisotropic turbulent (ordered) magnetic field strength, 
mainly along the azimuthal direction. This 
leads to a significant temporary increase of the 
total magnetic energy during the encounter; the representation of an increase in
star formation rate has an additional strong effect. In contrast to 
expectations, the large-scale (regular) magnetic field structure is not 
significantly destroyed by the encounter. It may be somewhat weakened for 
a relatively short period, and its direction after the encounter
may be reversed.}
{The encounter causes enhanced total and polarized emission without increase 
in the regular magnetic field strength.
The increase in synchrotron emission caused by the large-scale streaming
can be comparable to the effect 
of the increase in the star formation rate, depending on the choice of parameters.
The effects of the encounter on the total magnetic field energy last 
only slightly longer than the duration of the encounter (ca. 1~Gyr). 
However, a            
long-lasting field reversal of the regular magnetic field may result.}


\titlerunning{Magnetic fields enhancement arising from galactic encounters}
\authorrunning{Moss et al.}

\maketitle

\section{Introduction}
\label{int}

Galactic encounters are spectacular phenomena that are usually marked by
a substantial increase in synchrotron
emission of the interacting galaxy when compared with the typical emission from
similar non-interacting galaxies. 
The conventional interpretation associates this increase with an increase in
the star formation rate (SFR) and in the strength of turbulent magnetic fields resulting from 
the encounter (Schleicher \& Beck 2013). 
The radio--far infrared relation
also holds for interacting galaxies (e.g. Ivison et al. 2010). 
The polarized radio emission, signature of compressed
turbulent fields, is also known to be enhanced during the interaction (Vollmer et al. 2013).
In contrast, the regular magnetic field, which  is generated by the mean-field
dynamo and contributes to the synchrotron radiation,
is usually believed to decrease as a result of the encounter because the
interaction disturbs one of the drivers of galactic dynamos, namely the
rotation curve of the interacting galaxy. Of course, this naive
expectation requires verification.
In its full extent this would be a far from straightforward undertaking.
Indeed, a detailed modelling of the process
through an encounter would require modelling the magnetic fields and hydrodynamics
of both galaxies as well as the interaction itself.  This does not appear feasible
in the near future. There is  however a
simplification which isolates the features of the interactions that appear
salient for the evolution of the regular galactic
magnetic field, and can allow its evolution to be followed through an encounter event,
at least at an exploratory level.

Models for mean-field galactic dynamos have become increasingly more detailed
(and so arguably more realistic). However, even with the very significant
increase in computer resources that have become  available over the last
20 years or so, severe approximations remain necessary.
This is likely to remain true for the foreseeable future.
Broadly speaking, models split into two groups. In one group, attempts are made
to model in some detail processes in the ISM, including cosmic ray transport
and large-scale dynamics. Direct Numerical Simulation (DNS)
in ''boxes" can be used to provide estimates of transport coefficients 
(see e.g. Gressel et al. 2008, Brandenburg et al. 2008, Siejkowski et al. 2010). 
Models in the other group are in some ways less ambitious,
using a simpler and more direct mean field formulation, often with rather
{\it ad hoc} expressions for transport coefficients.
Brandenburg (2014) gives an up-to-date review.
 Of course there is considerable overlap of these approaches,
e.g. Hanasz, Woltanski \& Kowalik (2009).
The latter type of model is much less demanding of computing resources,
and readily allows extensive exploration of parameter space, and also the study
of phenomena such as large-scale gas streaming in barred galaxies
(e.g. Moss et al. 1998, 2001, 2007; Kulpa-Dybel et al. 2011; Kim \& Stone 2012), and the effects of spiral
arms on dynamo action (Shukurov 1998; Moss 1998; Chamandy et al. 2013a,b;
Moss et al. 2013).
There are also many studies of the effects of externally driven gas flows on
the gas content of galaxies (e.g. Vollmer, Braine \& Soida 2012 and references therein).
Some of these also solve the passive induction equation, but do not include dynamo action,
which is the main issue addressed in the current paper.

Our intention here is to revisit the effects of a galaxy-galaxy encounter,
which generates large-scale non-circular velocities, on large-scale dynamo action
in the larger galaxy.
This problem seems to have been first studied by Moss et al. (1993) and
Moss (1996). In the first of these papers,
non-circular velocities taken from a dynamical model
of the encounter between M~81 and its satellite NGC~3077  by
Thomasson \& Donner (1993)
were introduced into a (necessarily) rather low-resolution and
crude 3D dynamo model, and in the second a thin disc approximation was used.
Moss et al. (1992) used a static velocity field corresponding to the
instant of closest approach, whereas Moss (1997) used the fully time dependent
velocity field.
Although the dynamical model of the interaction is necessarily of low resolution
by contemporary standards, we feel it is adequate to investigate some
generic effects, and so we introduce the fully time dependent velocity field
from Thomasson \& Donner (1993) into the dynamo model recently proposed by
Moss et al. (2012). In this ''hybrid" dynamo model, a crude representation
of the effects of supernovae driven star formation regions in injecting
smaller scale magnetic field into the ISM are included explicitly in a thin disc
dynamo model. 
The resulting global magnetic fields display a number of novel features
for mean field dynamos, including disorder on the scale of a kiloparsec or
less, and small- and large-scale field reversals. We were particularly
interested in seeing what additional effects might result, arising from
the interaction of the non-circular velocities with the effects of the magnetic
field injections.
This sort of study has connections with cosmic ray driven dynamos of
e.g. Siejkowski et al. (2014).

 We first investigate the effects of the large-scale velocity field
that is generated by the encounter, on the galactic magnetic field,
and then tentatively explore the outcome of a parametrization of an
associated increase in the star formation rate. A problem here
is that the relation between the star formation rate and small-scale dynamo
action is not completely resolved. We have not
attempted any calibration of the novel
effects with any standard model which quantifies the effect of an
interaction on dynamo action via star formation,  and simply
represent this by a plausible parametrization.
We stress that this point deserves further attention;
however, such a study is obviously outside of the scope of this paper.

We describe our model in Section~\ref{model} including a brief recapitulation of 
the galactic dynamo model of Moss et al. (2012). Our main
result is that the interaction can significantly enhance synchrotron radiation emission,
by its effects both on the small-scale and regular fields. 
Our detailed results are presented in Sect.~\ref{results} followed by discussion and conclusions
in Sect.~\ref{disc}.

\section{The model}
\label{model}

\subsection{The dynamo setup}
The model is the thin disc model (''no-$z$" approximation)
 described in Moss et al. (2012), with the addition of advection of magnetic field by
the non-circular velocities.
 The model of Moss et al. (2012) has the novel feature that small-scale
field is continually injected at discrete locations, to simulate the
effects of star forming regions in introducing small-scale field into the ISM.
Briefly, we add random fields $B_{\rm inj}=B_{\rm inj0}f(r,t)$
at $n=250$ randomly chosen discrete locations
with re-randomization 
(i.e. changing the location of the injection sites and the
distribution of field strengths over them by choosing a new independent set of random numbers)
 at intervals
$dt_{\rm inj}\approx 10$ Myr.
We note that the no-$z$ approximation implicitly preserves
the solenoidality condition $\nabla\cdot {\bf B}=0$ for
both the dynamo generated and injected fields.
Full details are given in Moss et al. (2012).
We take a flat disc, while noting that there
is currently some uncertainty about whether galactic discs are substantially
flat or flared (cf. Lazio \& Cordes 1998); we note that further
investigation of this point  is needed, but it does not appear that our results
are very sensitive to this assumption. Our approach is consistent with that of 
Thomasson \& Donner (1993). 
The HI disc of the Milky Way does flare, but it is unclear whether the
ionized gas disc does so, and the observational data for external galaxies
are inconclusive.
In most of the simulations to be discussed, we took the conventional
alpha coefficient $\alpha$ and turbulent diffusivity $\eta$ to be uniform
throughout the disc; we intended a generic study with assumptions that
were as simple as possible. We did run models with
$\alpha(r)\propto\Omega(r)$ for comparison.

The code was implemented first on a Cartesian grid with $497\times 497$ points,
equally spaced, extended to just beyond the galactic radius,
taken as $R=15$ kpc.
In this  outer region beyond 15~kpc, there is no alpha-effect and the diffusivity retains
its global value.
This enables satisfactory treatment of the boundary conditions, see
Moss et al. (2012).
This grid was used until the interaction began, a statistically steady state
having been attained.
The computational grid was extended further after the interaction began,
to $617\times 617$
points, with the same mesh size as before, allowing a larger
exterior region to be included,  see Sect.~\ref{extvel} below.

We use a nondimensional time unit, $h^2/\eta$. With
$\eta=10^{26}$ cm$^2$ s$^{-1}$ (used in all but one model presented) and h=$500$ pc, this unit is approximately
$0.78$ Gyr. The timestep is fixed at approximately $0.04$ Myr.

\subsection{The velocity field}
\label{vel}
We adopted the velocity field for the interaction of M~81 and NGC~3077
of Thomasson \& Donner (1993), with some minor modifications implemented
for computational convenience. We took for the basic rotation curve the
(purely axisymmetric and azimuthal) velocities from the dynamical model
immediately  before the interaction began.
This had rather unsatisfactory behaviour near the galactic centre
where the velocities remained approximately constant until very near the centre.
Thus it was merged smoothly in the inner regions with a Brandt curve.
At subsequent times through the interaction (which lasted about $0.9$ Gyr),
the non-circular velocities were as generated by the Thomasson \& Donner code;
the perturbations to the $m=0$ azimuthal velocities at a given time
were calculated by subtracting the raw  $m=0$ azimuthal velocities before the
interaction from the current values. Fourier modes $m=1, 2, 3$ were included
from the Thomasson \& Donner data.
The interaction generates non-zero velocities for about 0.9~Gyr.

\subsection{Choice of parameters}
\label{param}
The large-scale velocities in the system are given (in km s$^{-1}$)
by the dynamical model,
so once the values of the turbulent diffusivity and disc thickness are chosen,
the conventional dynamo parameter $R_\omega$ and the magnetic Reynolds number
of the non-circular motions $Rm$ are fixed. For reference, with
$\eta=10^{26}$ cm s$^{-1}$ and $h=500$ pc, then $R_\omega=0.75, Rm=1.5$, and these
values were used for the bulk of the simulations.
For illustrative purposes we ran some
models with other values of these parameters. The value of the parameter
$R_\alpha$ is more uncertain: we took $R_\alpha=\alpha_0h\eta^{-1}=1$ as
 our standard value for the flat disc models.
We note that we assume that $\alpha$ and $\eta$ are uniform through the disc.

\begin{table*}
\caption{The salient models. The parameters are those applicable 
during the interaction - $Rm$ is zero outside of this interval. 
Y in Col.~6 indicates a model with 
$\alpha\propto \Omega$, else N, see \protect Sect.~\ref{param}. 
The last column indicates whether the injection magnitude $B_{\rm inj0}$ 
(proxy for SFR) was arbitrarily increased
by the encounter, as Eq.~(2), and gives the value of $q_{\rm I}$.
$q_{\rm I}=0$ indicates no enhancement. $\eta$ is given in cm$^2$ s$^{-1}$.}
\begin{tabular}{|c|c|c|c|c|c|c|c|}
\hline
Model & $\eta$  & $R_\alpha$  & $R_\omega$ & $R_{\rm m}$ & $\alpha \propto \Omega$  & $B_{\rm inj0}$ & $q_{\rm I}$ -- Eq.~(2) \\
\hline
9 &    $10^{26}$ & 1.0 & 0.75 & 1.5 & N & 0 & 0\\
15 &    $10^{26}$ & 1.0 & 0.75 & 1.5 & N & 4.0 & 0\\
16 &    $10^{26}$ & 1.0 & 0.75 & 1.5 & N & 8.0 & 0\\
17 &   $2\cdot 10^{26}$ & 1.0 & 0.375 & 0.75 & N & 8 & 0\\
101 &    $10^{26}$ & 3.0 & 0.75 & 1.5 & Y & 8.0 & 0\\
102 &    $10^{26}$ & 6.0 & 0.75 & 1.5 & N & 8.0 & 0\\
203 &    $10^{26}$ & 1.0 & 0.75 & 1.5 & N & 8.0 & 2.0\\
204 &    $10^{26}$ & 1.0 & 0.75 & 1.5 & N & 8.0 & 9.0\\
\hline
\end{tabular}
\label{summary}
\end{table*}

\subsection{Extension of the model beyond the galactic radius}
\label{extvel}
We found that when we set the boundary of the galactic disc at $r=R=15$ kpc,
which is the extent of the dynamical model and thus of the given velocities,
then anomalous field gradients were generated near the boundary. 
For this reason we
extended the velocities into the region $r>R$, reducing to very small values
by $r=1.35R$.
This is certainly unphysical, in that we expect rotation curves to
remain more-or-less flat until quite large radii. However, the interaction model
does not provide data beyond $r=R=15$ kpc, and this exterior region
is only included to provide a satisfactory treatment of the outer
parts of the disc; i.e. we do not try to represent accurately the region $r>R$.
In $r>R$ diffusion acts, but there is no alpha-effect.
The dynamo model of Moss et al. (2012) is already embedded in a surrounding
passive region, so this is a minor modification which is found to
give more satisfactory behaviour near $r=R$, and allows fields to be advected
weakly beyond radius $R$.

\subsection{Parametrization of the connection between the star formation rate and the dynamo governing parameters}
\label{SFR}
In order to compare the effect of changes to the SFR caused by the interaction on the dynamo action we have to choose
a parametrization of dynamo governing parameters which includes the star
formation rate. Here we face the problem that
such parametrization is little discussed in the current literature (see e.g. Mikhailov et al. 2012).
However,
this is an important problem which needs to be addressed
separately.
In order to make some progress we adopt a parametrization which looks
at least plausible at first sight, i.e. we anticipate
a simple quasi-linear relation between the 
SFR and the energy input into turbulence. This
seems quite natural because a higher SFR means more supernovae and hence
more energy input. As each supernova is a singular event, the energy
input should simply add linearly. However, the question is whether
the energy fractions going into turbulence, cosmic rays, and gas heating depend
on the overall level of star formation or on the density of the
surrounding medium. A higher SFR is expected in denser galaxies where the
typical Mach numbers of supernova shocks should be higher.
At the moment it remains unclear to what extent this
affects the energy fractions.
Additional uncertainties include whether a higher SFR is associated
only with denser concentrations of SN, or larger star forming regions,
or a combination.
However, in any case it is to be expected that an increase in SFR will
produce increased small-scale dynamo action.  We represent this by an
increase in the parameter $B_{\rm inj0}$, but could also consider increasing the
size of the injection sites, or their number. Furthermore, we do not know the relation between
energy deposited and the resulting small-scale
dynamo field strength (see however Geng et al. 2012, A. Beck et al. 2012, which support
the general view that something like equipartition will occcur).

We stress that the point deserves clarification;
however, we restrict ourselves to this simple
parametrization.

\section{Results}
\label{results}

\subsection{The standard model}
\label{stand}
The more relevant models of those computed are summarized
in Table~\ref{summary}.

We first discuss a model with the standard parameters described in
Sect.~\ref{param}, that is $R_\alpha=1, R_\omega=0.75, Rm=1.5$, $f(r,t)=1$.
Our computational procedure was to run the code until dimensionless time $\tau=17$,
corresponding to an age of approximately 13 Gyr (''now",
cf Moss et al. 2012). As the field has long been settled
into a statistically steady state
by this time, the choice $\tau=17$ is not particularly significant
in the current context and, for our purposes, it is the time from the start
of the interaction at $\tau=17$ that is relevant.
The non-circular velocities are then introduced, and the code  is run for a
further dimensionless time of about 4 units. As the interaction lasts
for a little over 1 dimensionless time unit (ca 0.9 Gyr), by the time the
simulation ends the field is found to approach a statistically steady state
again. These models are basically similar to those of Moss et al. (2012),
except for the different rotation profile and larger galactic radius
(here $R=15$ kpc compared to $10$ kpc).

\begin{figure*}
\begin{center}
\begin{tabular}{lll}
(a)  \includegraphics[width=0.3\textwidth]{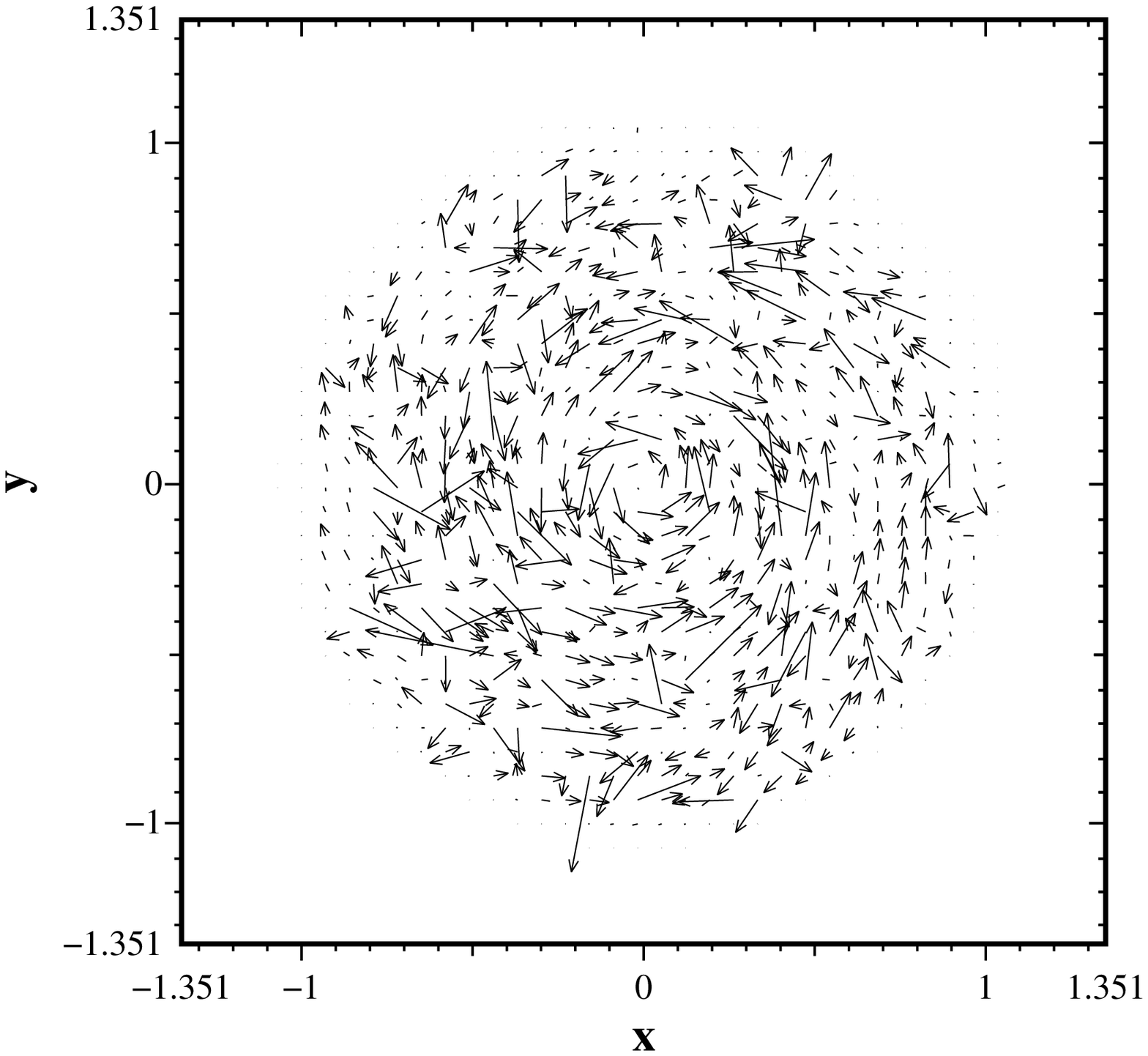} &
\includegraphics[width=0.3\textwidth]{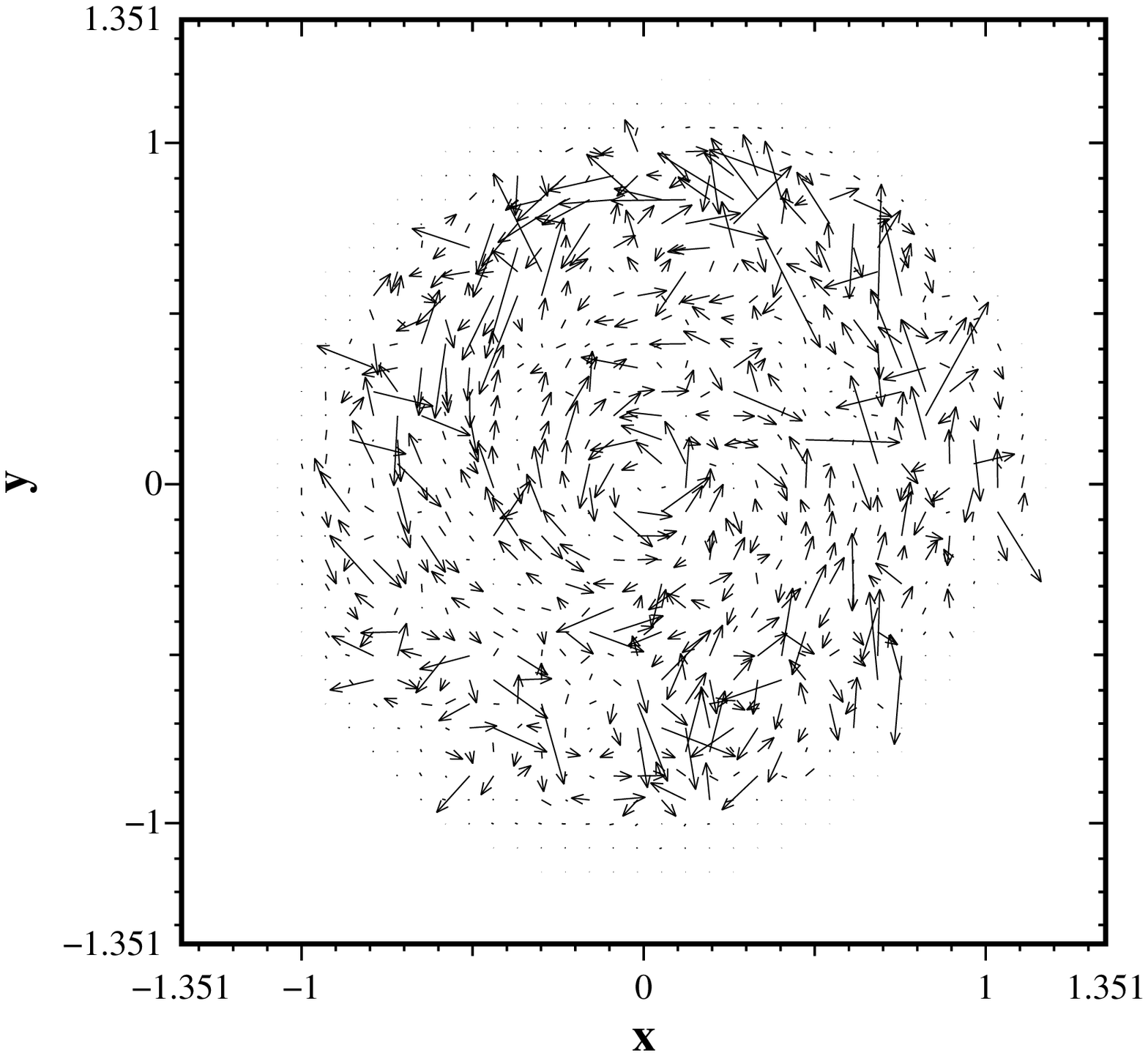}&
 \includegraphics[width=0.3\textwidth]{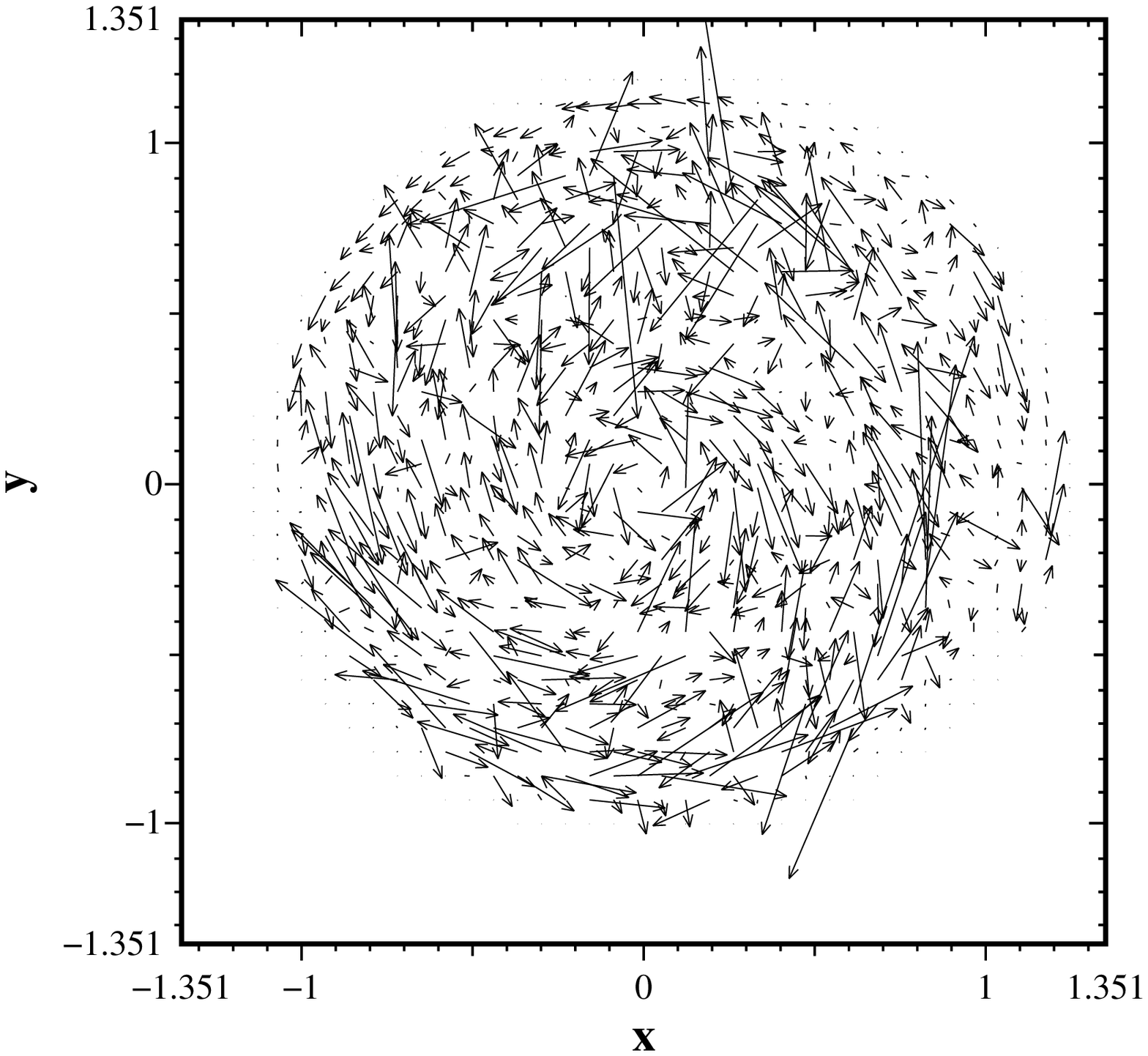} \\
(b) \includegraphics[width=0.3\textwidth]{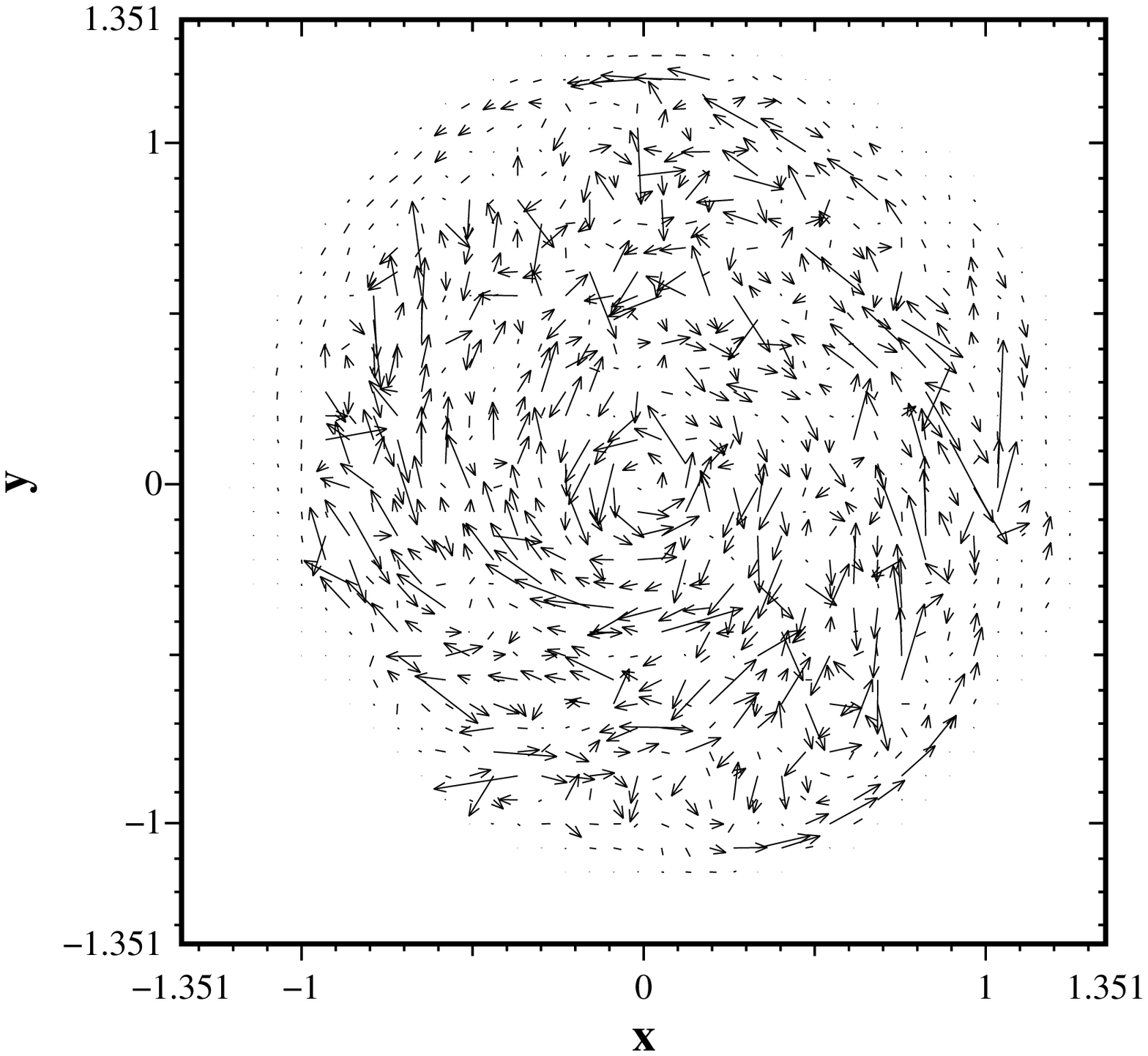}&
\includegraphics[width=0.3\textwidth]{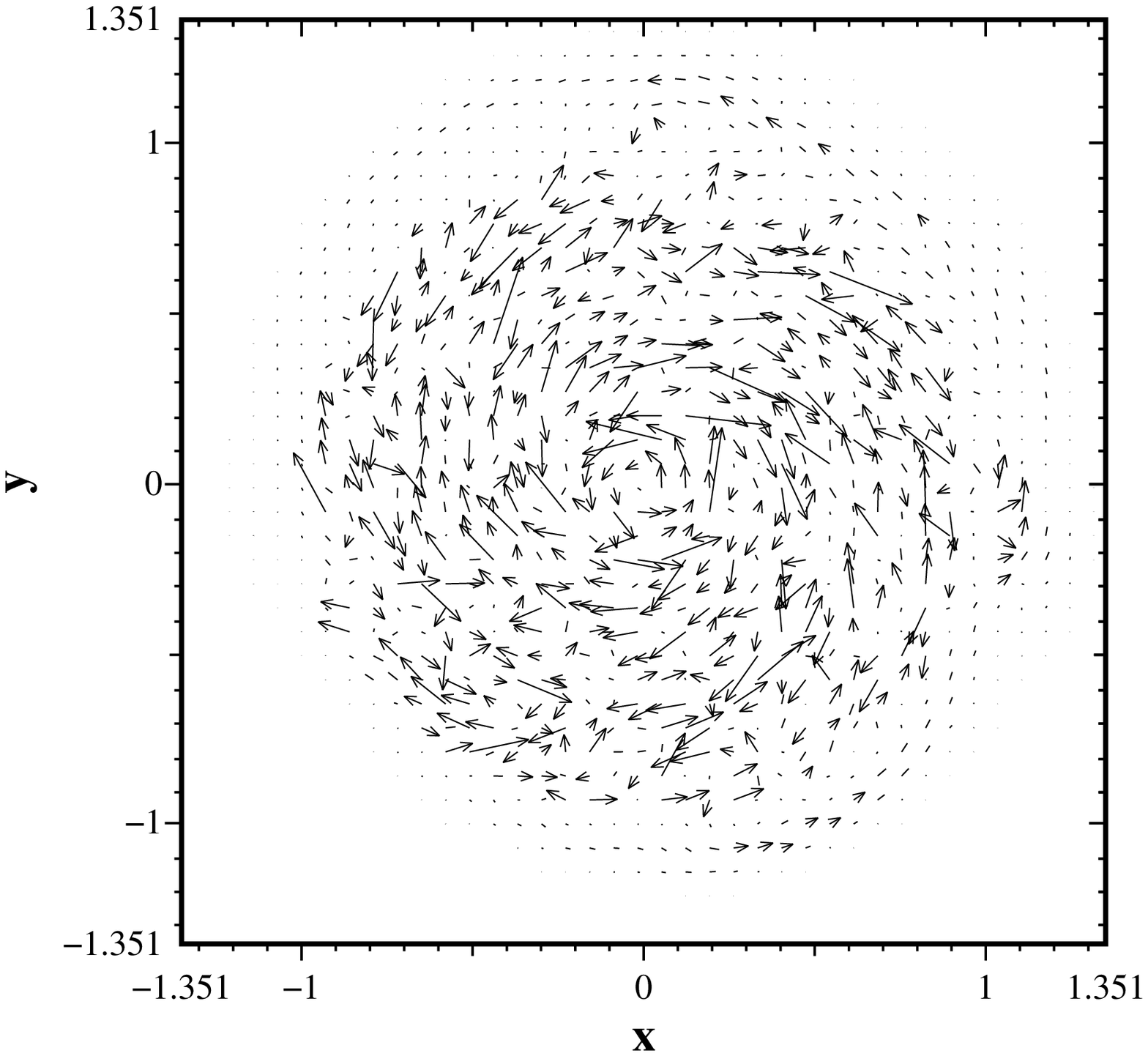} &
\includegraphics[width=0.3\textwidth]{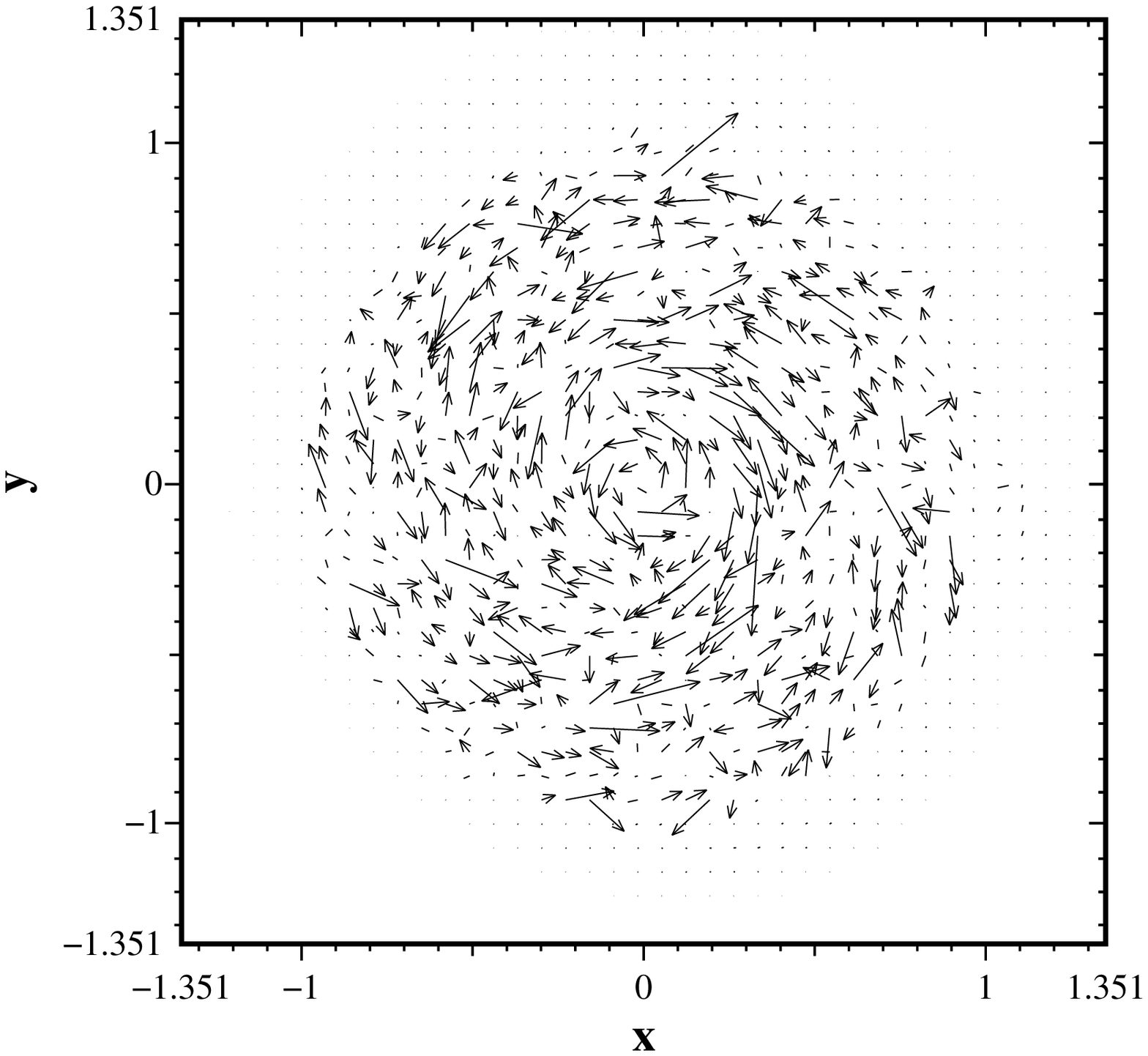}
\end{tabular}
\end{center}
\caption{Field vectors at times $t \approx 13.3, 13.65, 13.8$ Gyr (row (a)),
and $t\approx 14.0, 14.3, 14.7$ Gyr (row (b)), for
the canonical model 16.  The interaction extends over the interval $13.3 \le t \lesssim 14.2 $ Gyr. The vectors
give the magnetic field direction, and their lengths are
proportional to the magnetic field strengths.
}\label{modr16x}
\end{figure*}
\begin{figure}
\includegraphics[width=0.90\columnwidth]{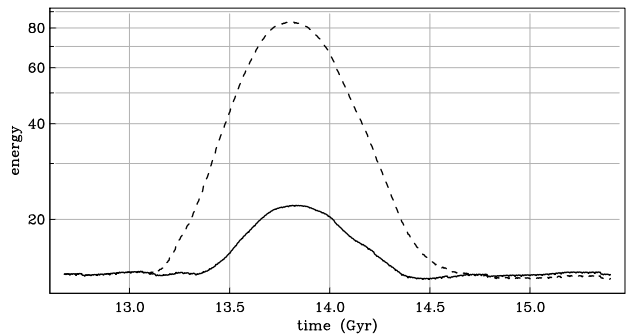}
\caption{Running time averages of total magnetic energy for Model 16 (solid)
and Model 203 (broken). The
interaction begins at $t\approx 13.3$ Gyr and ends at $t\approx 14.2 $ Gyr.}
\label{modr16x_tav}
\end{figure}

In Fig.~\ref{modr16x} we show snapshots of the magnetic field, from the
beginning to just after the end of the interaction.
We also show in Fig.~\ref{modr16x_tav} (solid curve) the time-averaged magnetic energy.
(Time averages are used to smooth the instantaneous effects of the
field injections,
and are taken with a sliding window $\Delta\tau=0.25$,
corresponding to physical time intervals of about $0.2$ Gyr, 
except for Model 17 where $\Delta\tau=0.50$.)

\begin{figure*}
\begin{center}
\begin{tabular}{lll}
(a)  \includegraphics[width=0.3\textwidth]{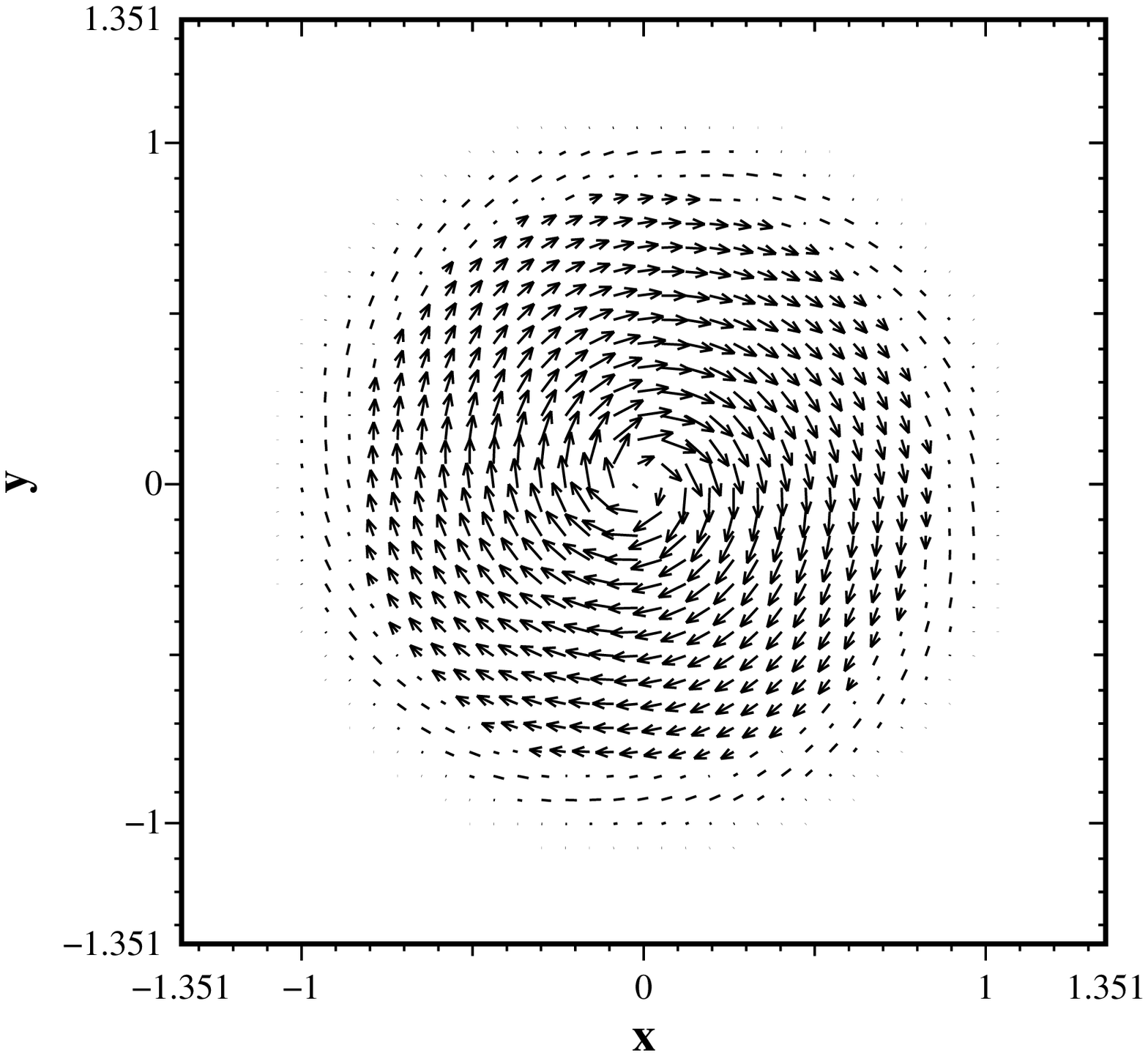} &
\includegraphics[width=0.3\textwidth]{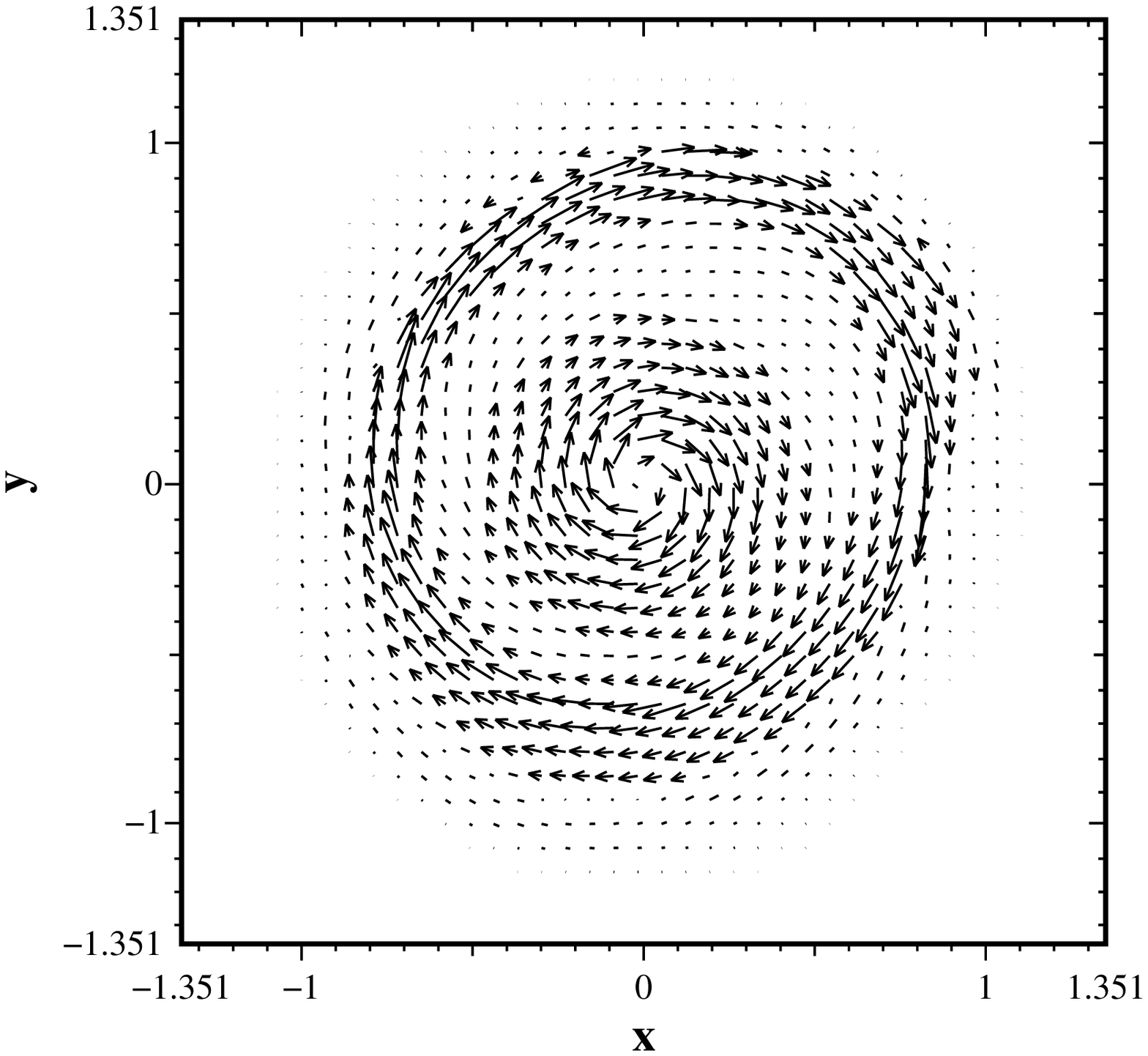}&
\includegraphics[width=0.3\textwidth]{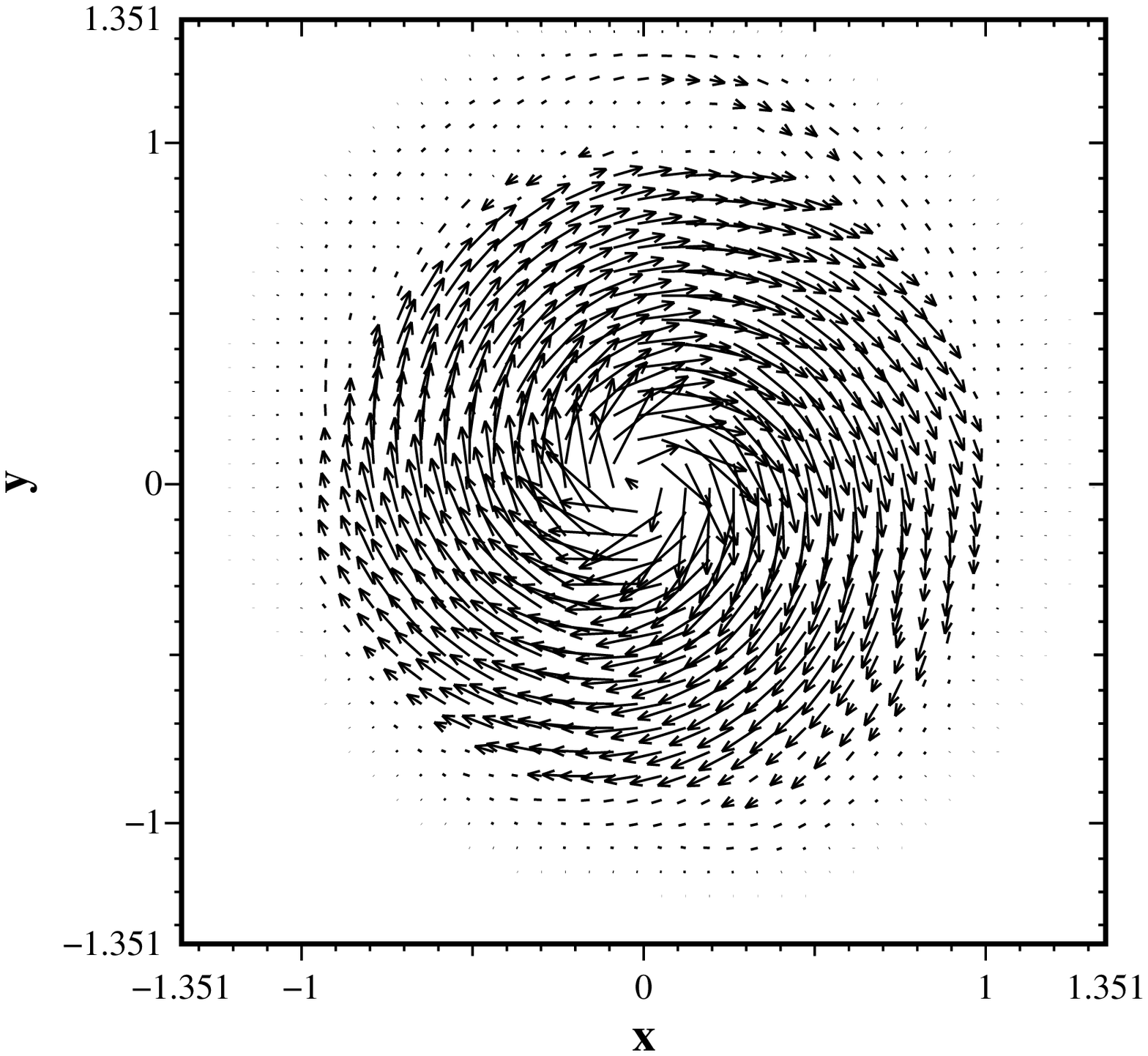} \\
(b) \includegraphics[width=0.3\textwidth]{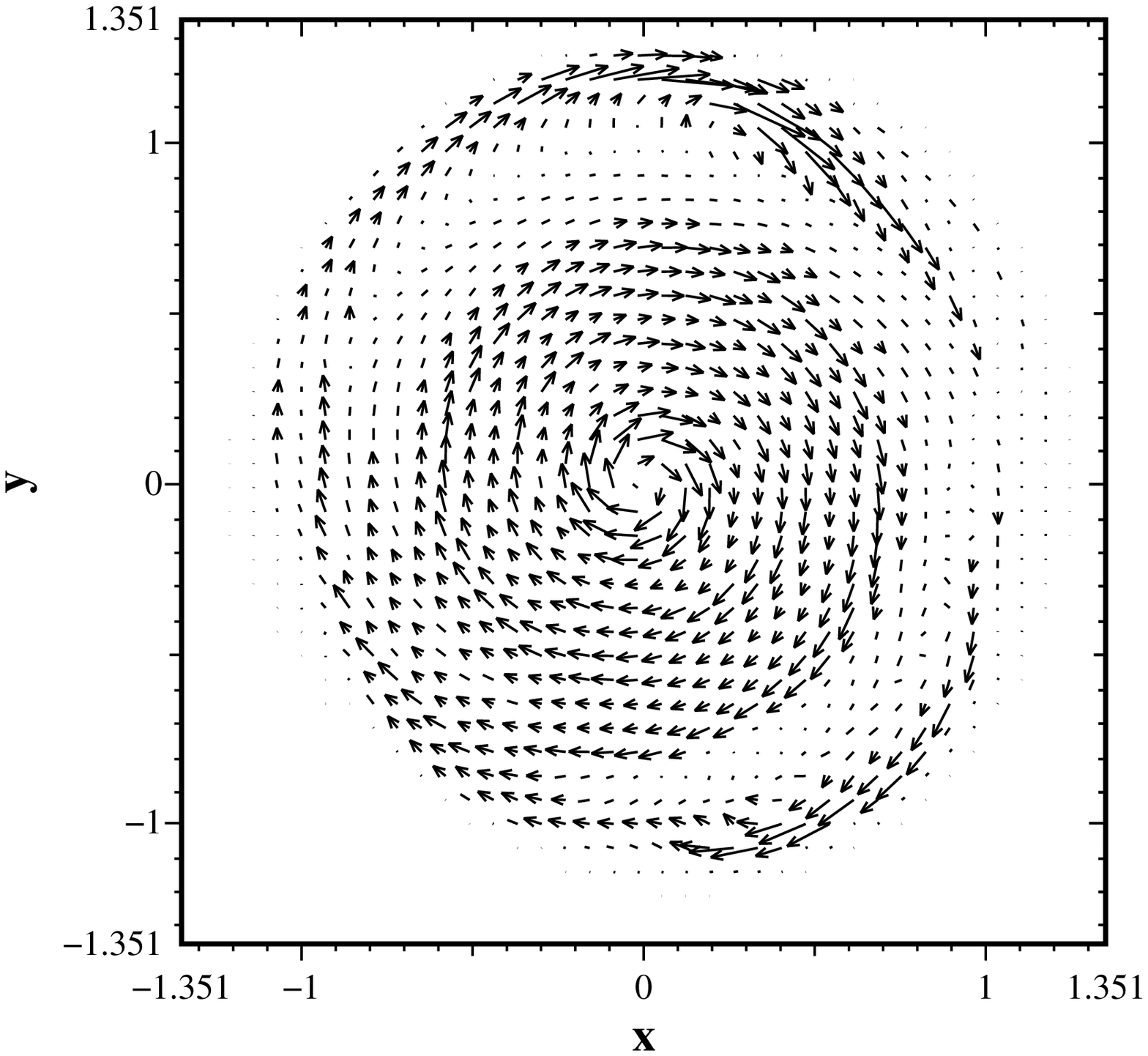}&
\includegraphics[width=0.3\textwidth]{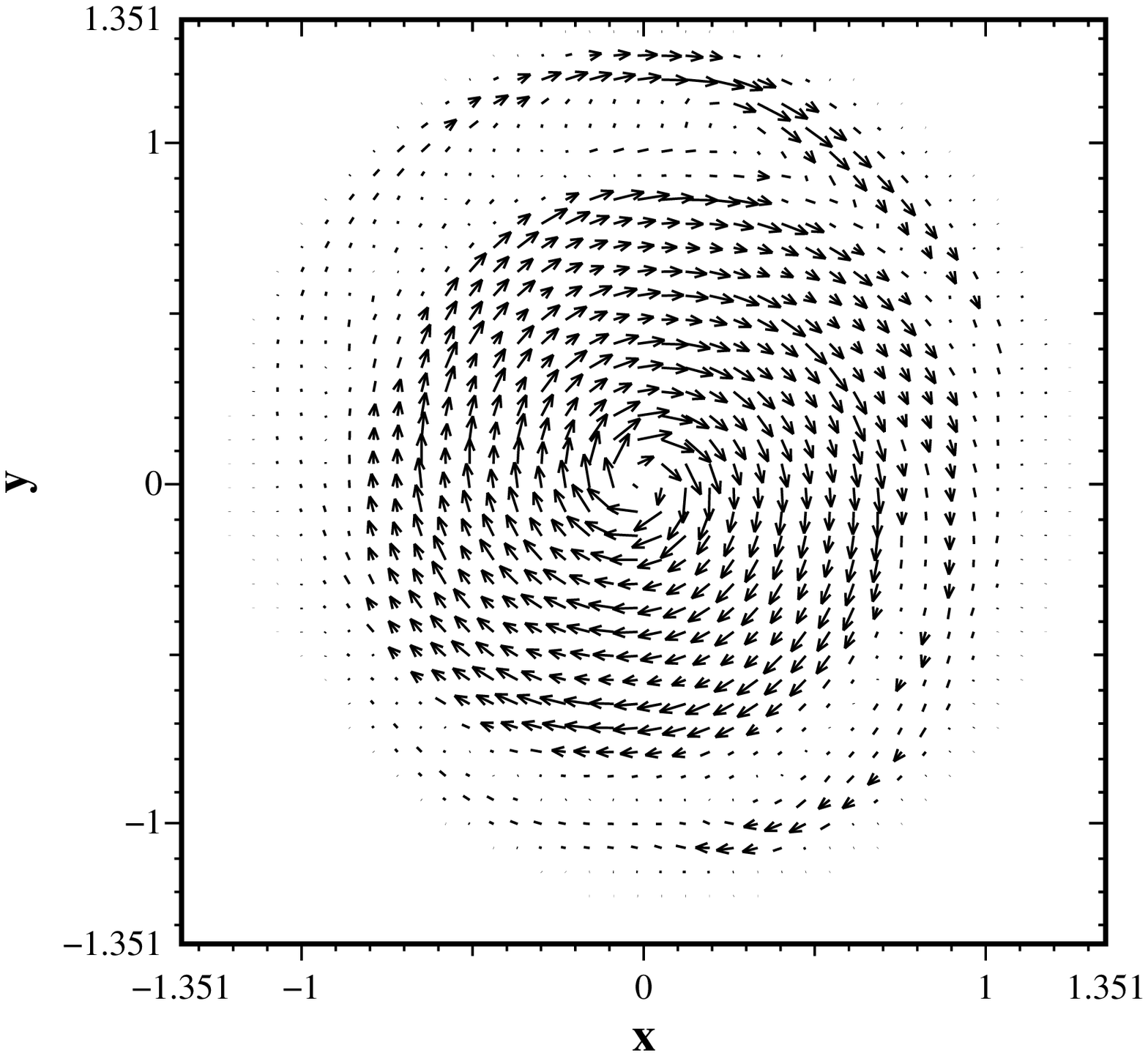} &
\includegraphics[width=0.3\textwidth]{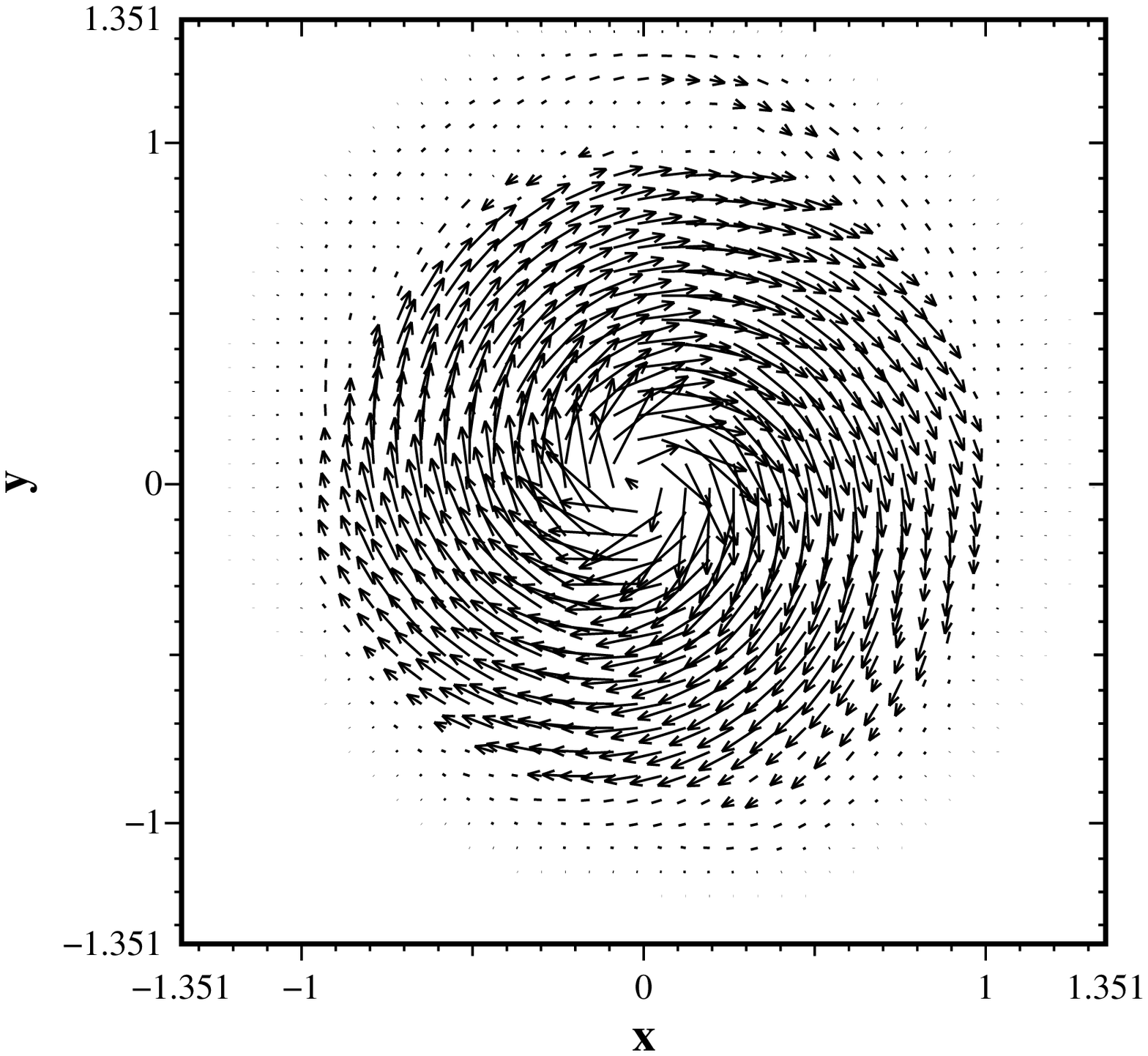}
\end{tabular}
\end{center}
\caption{Field vectors at dimensionless times $t\approx 13.3, 13.65, 13.8$ Gyr (row (a)),
$t=14.0, 14.3, 14.7$ Gyr (row (b)),  for
the conventional Model 9 (without field injections) . The vectors
give the magnetic field direction, and their lengths are
proportional to the magnetic field strengths.
}\label{modr9x}
\end{figure*}

\begin{figure}
\includegraphics[width=0.90\columnwidth]{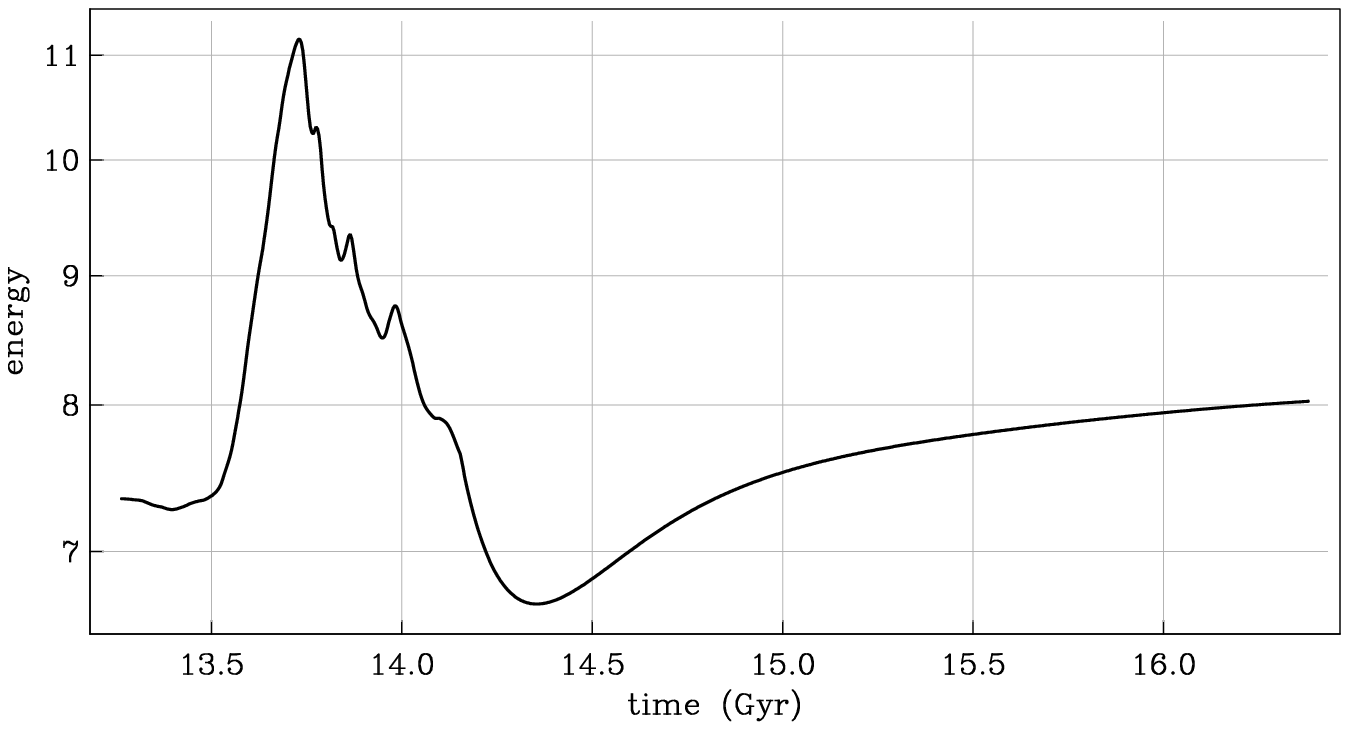}
\caption{Model 9,  total magnetic energy. The
interaction begins at $t\approx 13.3$ Gyr.}
\label{modr9x_t}
\end{figure}

To provide some orientation, we compare these results with a more conventional
dynamo model (Model 9) with no field injections. This model has the same
dynamo parameters as Model 16, and again was first run until time  $t\approx 13.3$ Gyr,
by which
time a steady state had long been established. At this time the interaction
began, with the same underlying rotation curve as Model 16. Snapshots of the
field configuration through and beyond the interaction are shown in
Fig.~\ref{modr9x}. In slight contrast to Model 16 (see Fig.~\ref{modr16x}), 
in the last panel ($t\approx 14.7$ Gyr)
there are still small signs of the interaction process in $r>R$,
but a steady state is
reached by time 15.6 Gyr. Interestingly, the stable reversal present in the
field before the interaction (first panel of Fig.~\ref{modr9x})
is absent from the
post-interaction field, and the magnetic energy
(shown in Fig.~\ref{modr9x_t}) does not return to its original value.

In both of these simulations, the effect of extending the velocity
field beyond the nominal boundary of the galaxy at $R=15$ kpc is visible,
dragging the magnetic field into the exterior region. This is reminiscent of the
extended field structures seen when modelling the effects of intergalactic
''winds", studied by Moss, Sokoloff \& Beck (2012).
If the velocities in $r>R$ are allowed to decrease more slowly with radius,
then these ''tails" extend farther in radius. Given the physical
uncertainties associated with these regions, we did not pursue this point.

\begin{figure*}
\begin{center}
\begin{tabular}{lll}
(a)  \includegraphics[width=0.3\textwidth]{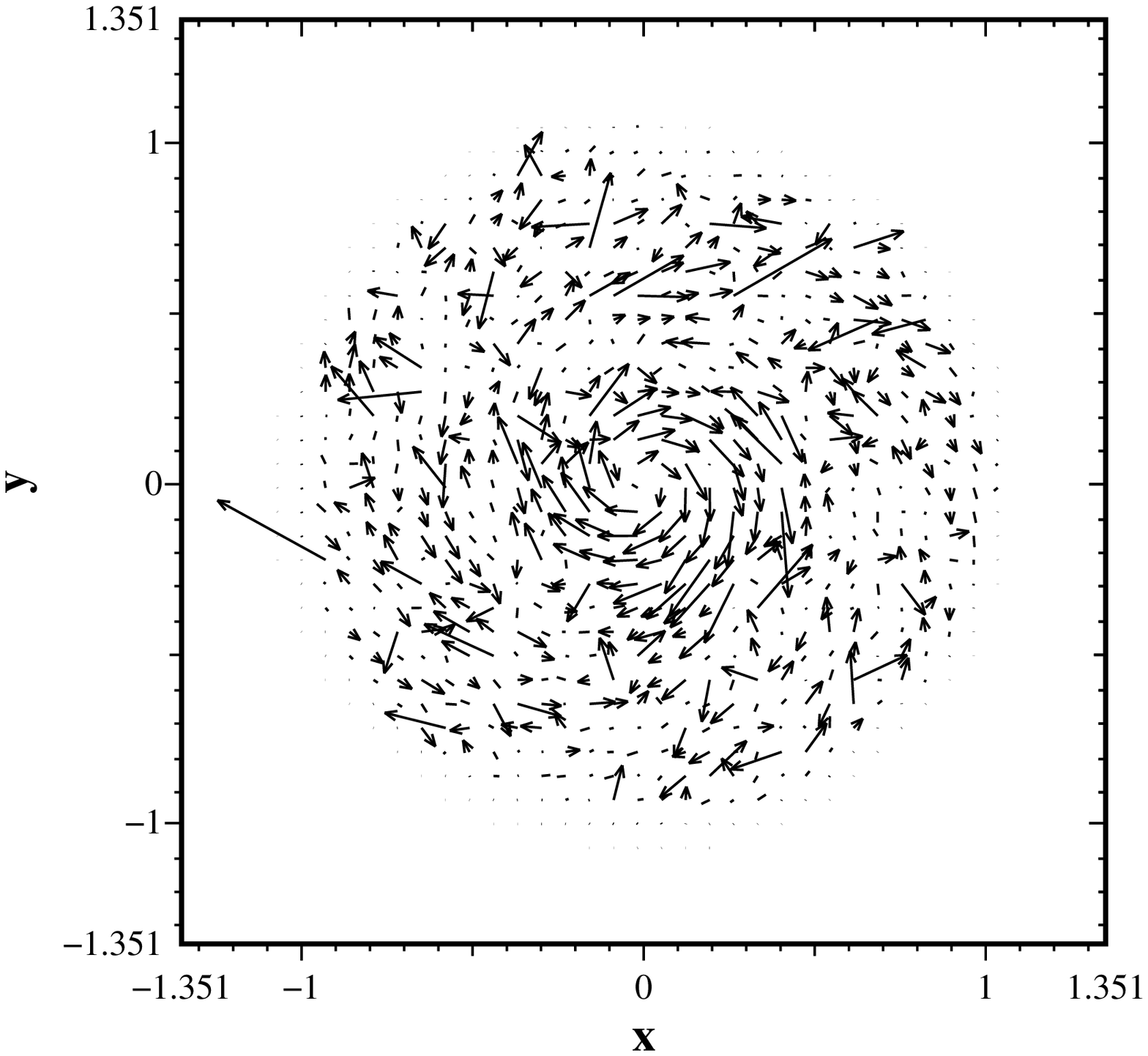} &
\includegraphics[width=0.3\textwidth]{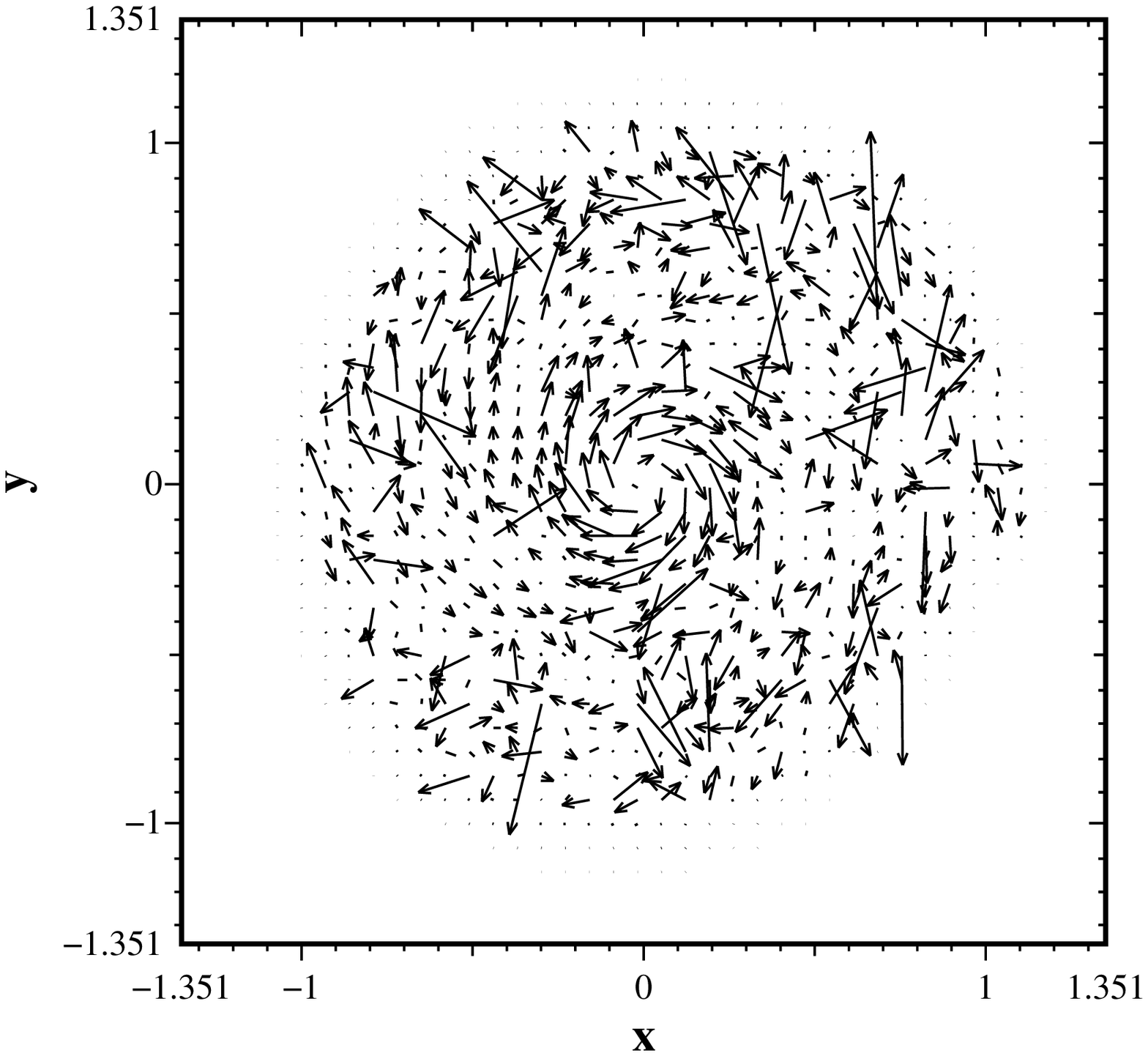}&
\includegraphics[width=0.3\textwidth]{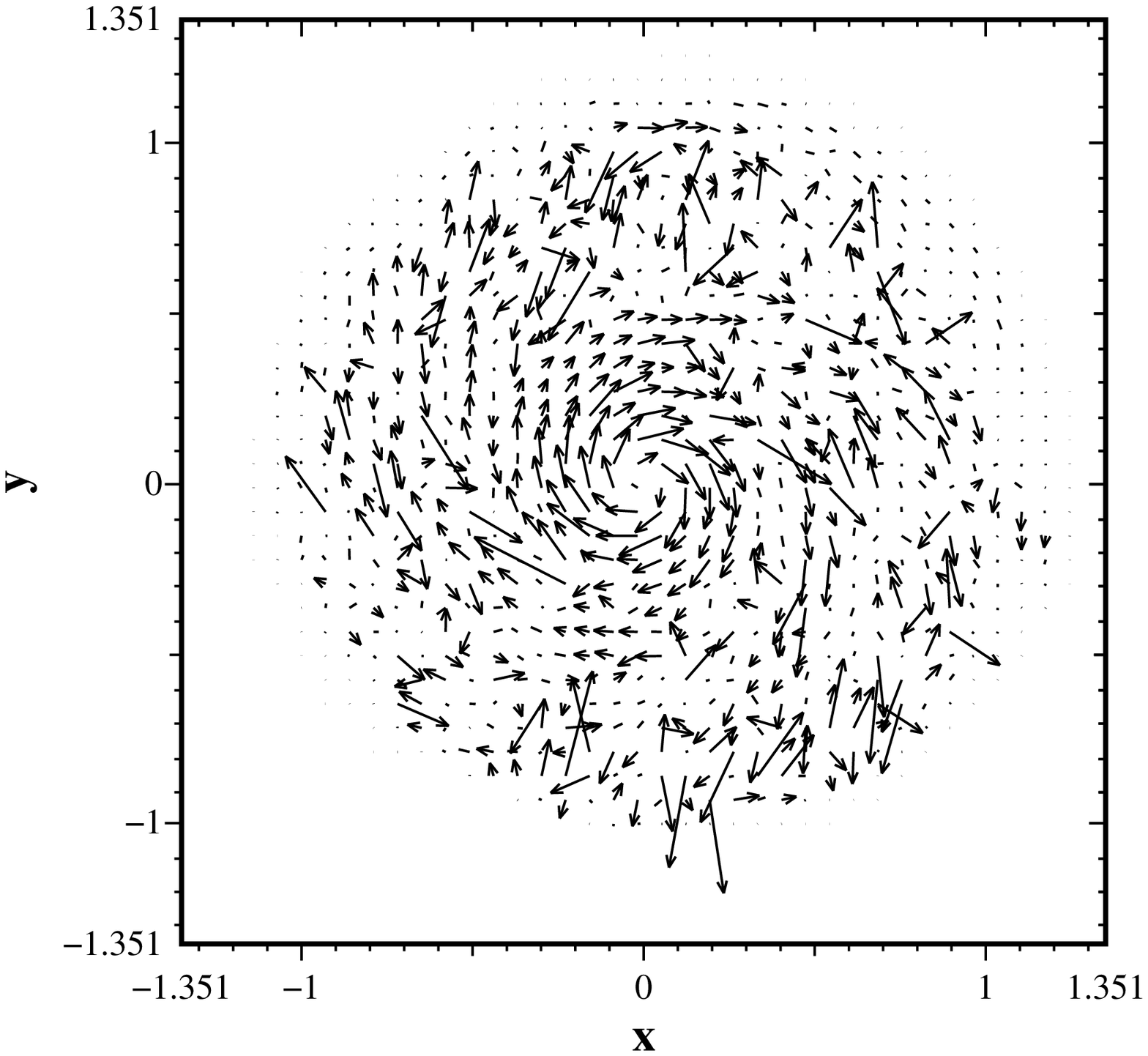} \\
(b)\includegraphics[width=0.3\textwidth]{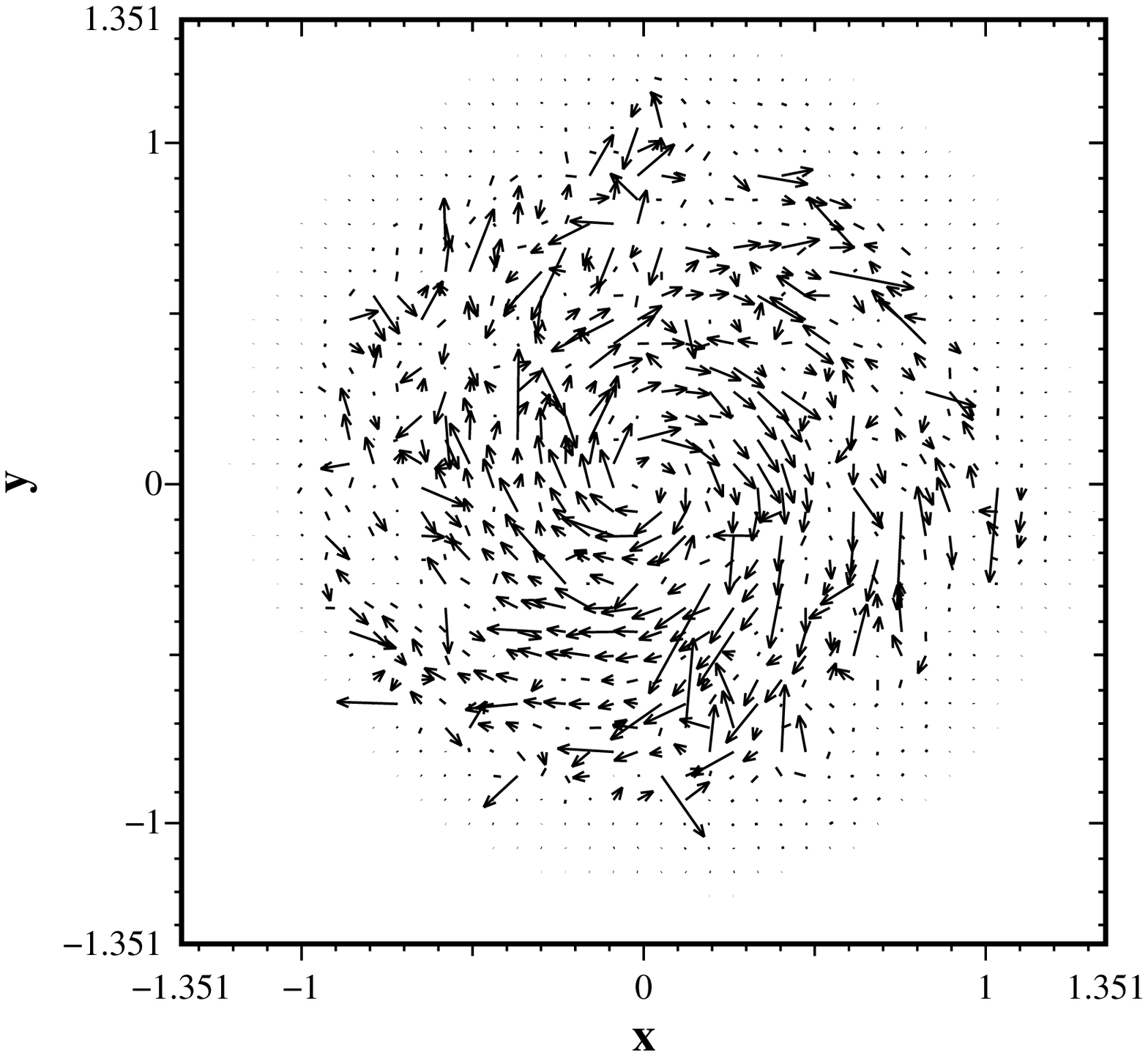}&
(c)  \includegraphics[width=0.3\textwidth]{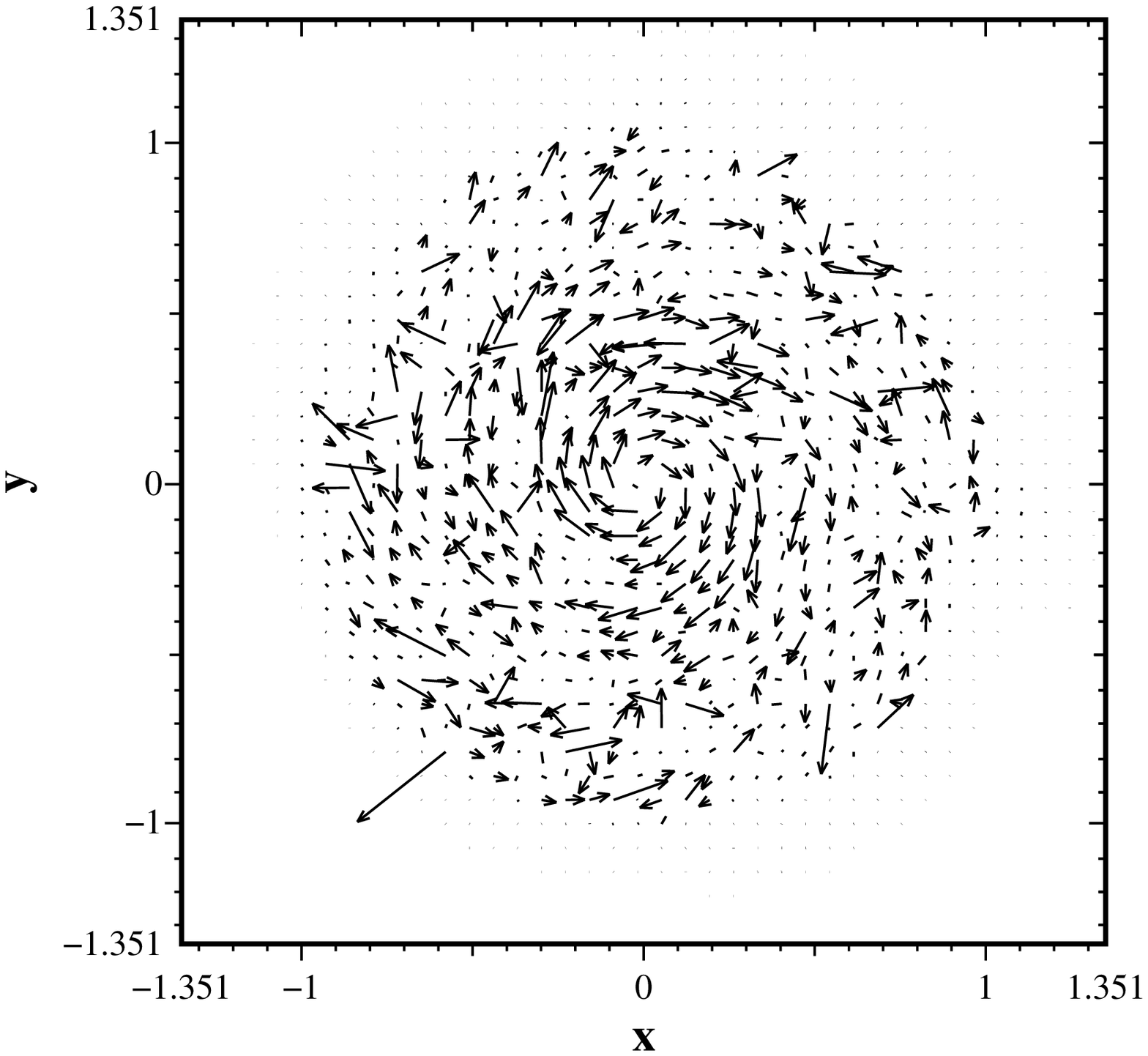}% &
\end{tabular}
\end{center}
\caption{Field vectors at dimensionless times $t\approx 6.6, 7.0, 7.3 $ Gyr (row (a)),
$t\approx 7.5,  7.8$ Gyr (row (b)), for
Model 17 (with $\eta=2 \cdot 10^{26}$ cm$^2$ s$^{-1}$). The vectors
give the magnetic field direction, and their lengths are
proportional to the magnetic field strengths.
The interaction occurs during $ 6.6\le t  \lesssim  7.5$ Gyr.
}\label{modr17x}
\end{figure*}

\begin{figure}
\includegraphics[width=0.90\columnwidth]{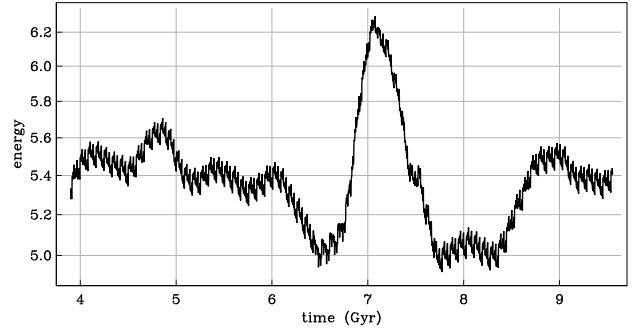}
\caption{Model 17, running time average of total magnetic energy. The
interaction begins at $t\approx 6.6$ Gyr and ends at $t\approx 7.5$ Gyr.}
\label{modr17x_tav}
\end{figure}

The value taken for the turbulent diffusivity, $\eta=10^{26}$ cm$^2$ s$^{-1}$,
is conventional but uncertain. Thus, it is desirable to investigate
the effects of changing the value of $\eta$ (and so the magnetic Reynolds numbers).
We found that the code could not handle the increase in $R_\omega$, $Rm$
resulting from a decrease in $\eta$ at an affordable resolution,
but we show in Figs.~\ref{modr17x} and \ref{modr17x_tav} the results of increasing
$\eta$ to $2 \cdot 10^{26}$ cm$^2$ s$^{-1}$.
As the time unit scales with $\eta^{-1}$, the values of $\tau$ for the panels
of Fig.~\ref{modr17x} differ substantially from those in the
earlier figures, but the time intervals from the onset of
the interaction are similar.
The smaller value of $R_\omega$ taken with an unchanged mean injected field strength
means that fluctuations are larger in Model 17 than in Model 16.
Also note that the local dynamo number is everywhere reduced by the increase in $\eta$.

\subsection{A model with $\alpha(r)\propto \Omega(r)$}
\label{omeg}

\begin{figure}
(a)\includegraphics[width=0.75\columnwidth]{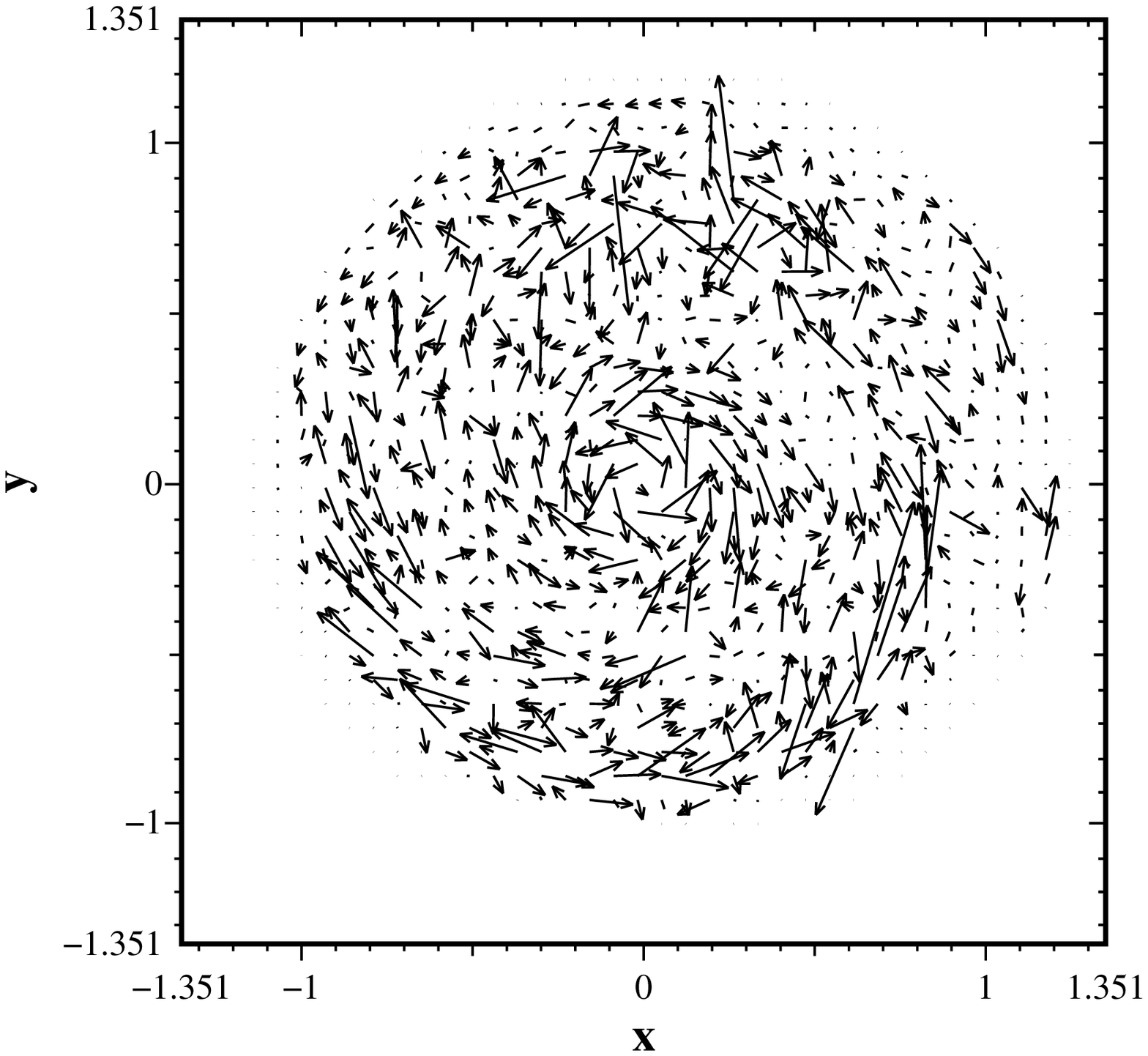}\\
(b)\includegraphics[width=0.75\columnwidth]{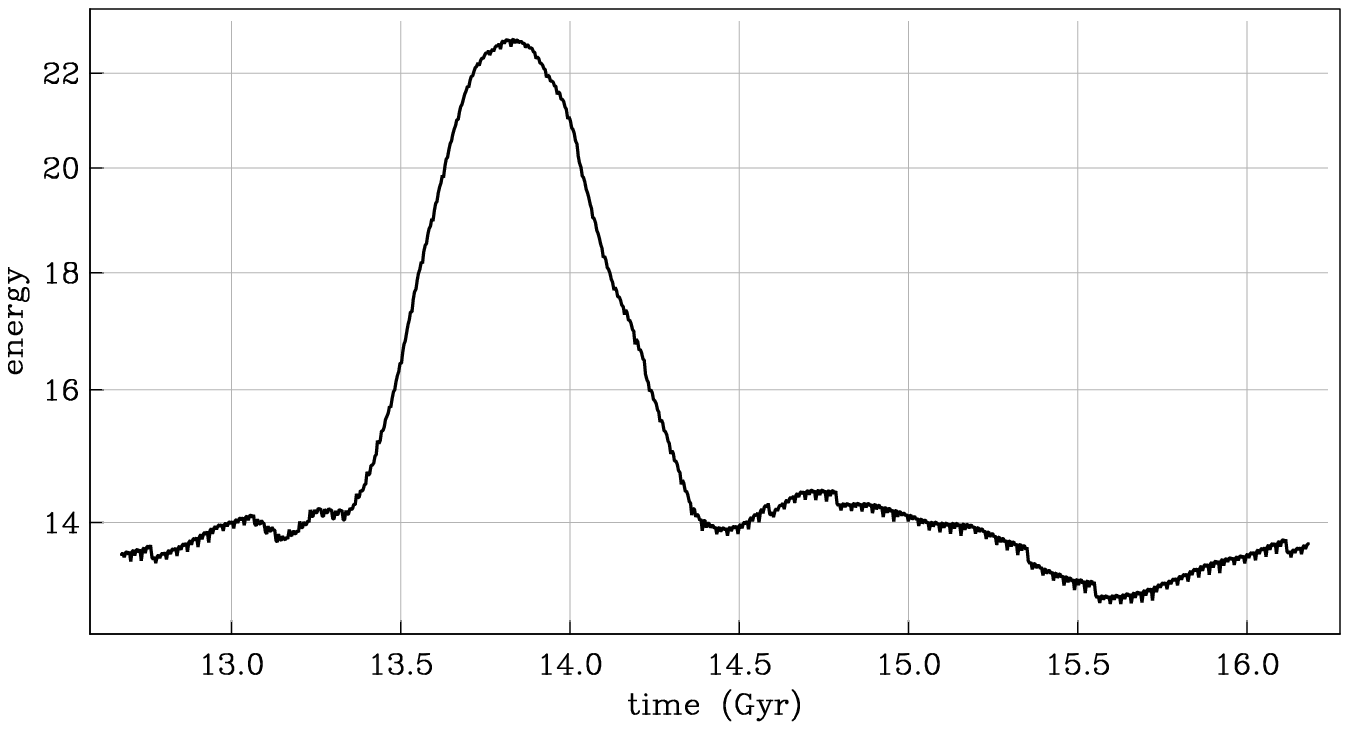}
\caption{Model 101 (with $\alpha\propto \Omega$); (a) field at $t\approx 13.85$ Gyr,
(b) running time average of total magnetic energy. The
interaction begins at $t\approx 13.3$ Gyr and ends at $t\approx 14.2 $ Gyr.}
\label{modr101x}
\end{figure}

In order to verify that our simple assumptions, in particular
that $\alpha$ is constant throughout the disc, were not affecting the results significantly,
we ran cases with $\alpha(r)\propto \Omega(r)$ (see
e.g. Ruzmaikin et al. 1988). The value of $R_\alpha$ was now defined
in terms of the central value $\alpha(0)$, and we ran several cases with
the parameters of Model 16, but $R_\alpha>1$ (to compensate for the decrease
in alpha with radius). We show in Fig.~\ref{modr101x} the field configuration
in mid-interaction at time $t=13.85$ Gyr, and also the running time-averaged global energy,
with $R_\alpha=3$.
Comparison with the third panels of Figs.~\ref{modr16x} and \ref{modr16x_tav}
suggest that there are few qualitative differences.

\subsection{Dynamo action versus star formation in an encounter event}

An enlargement of the total flux of synchrotron emission can also be connected
with the well known increase in the SFR caused by an encounter.
According to numerical simulations by Matteo et al. (2008), the increase in the SFR
is by a factor smaller than 5 in about $85 \% $ of all major galaxy interactions
 and mergers for low 
redshift galaxies. To estimate the relative influence of the increase in SFR compared
to that of the dynamo action and large-scale non-circular velocities alone,
we performed the following experiments.

We took the basic Model 16, and in one case multiplied the field injection rate
by the factor
\begin{equation}
f_{\rm I,1}=1+q_{\rm I}\frac{(\tau-\tau_{\rm beg})(\tau_{\rm end}-\tau)}{0.25(\tau_{\rm end}-\tau_{\rm beg})^2}
\label{enhanc1}
\end{equation}
between times $\tau_{\rm beg}$ and $\tau_{\rm end}$,
where $\tau_{\rm beg}, \tau_{\rm end}$ are the times of the
start and finish of the interaction (i.e. when non-circular velocities are non-zero).
This model attempts to simulate a global increase in SFR , which is
increased by a factor of $(1+ q_{\rm I})$
at mid-interaction.
Perhaps not very surprisingly, even with $q_{\rm I}=1$ there is a marked increase
(by a factor of about 5) in the global magnetic energy at mid-interaction.
%We show the evolution of the time-averaged global energy in Fig.~\ref{r201xtav.eps}.

In the second case we set
\begin{equation}
f_{\rm I,2}=1+q_{\rm I}\frac{(\tau-\tau_{\rm beg})(\tau_{\rm end}-\tau)}{0.25(\tau_{\rm end}-\tau_{\rm beg})^2})\exp{\left(-\frac{r}{R}\right)},
\label{enhanc2}
\end{equation}
simulating an increase in SFR
concentrated towards the central regions.
Again, the global energy at mid-interaction increases markedly compared
to the standard case, Model 16 (but by less than
with $f_{\rm I,1}$ given by Eq.~(\ref{enhanc1}) for the same $q_{\rm I}=1$).
Thus the Models discussed above in Sects.~\ref{stand} and \ref{omeg} effectively take $q_{\rm I}=0$.

The plots of magnetic field vectors do not show marked differences to those displayed
in Fig.~\ref{modr16x}, and so we do not show them.

\begin{figure}
(a)\includegraphics[width=0.80\columnwidth]{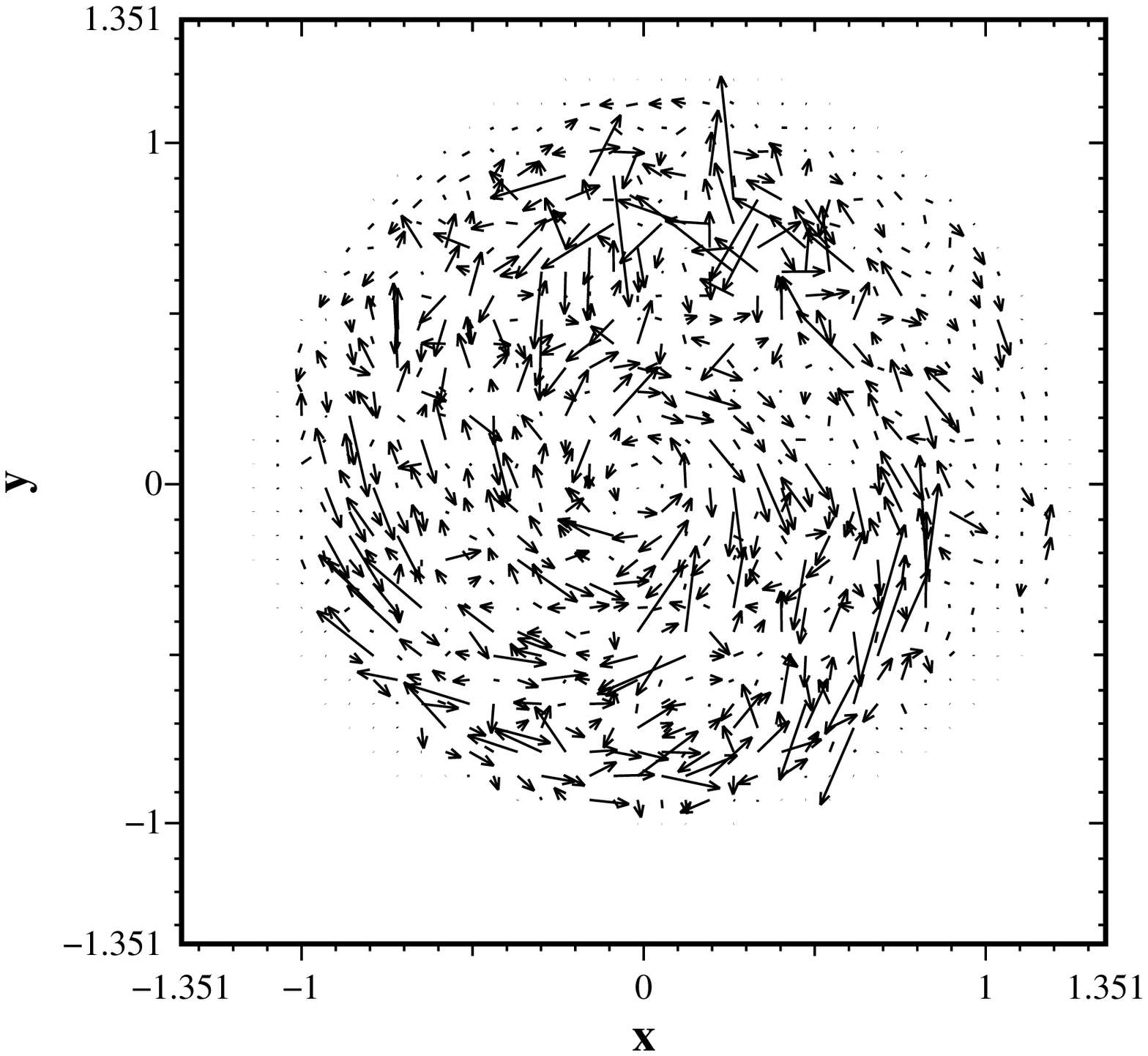}\\
(b)\includegraphics[width=0.80\columnwidth]{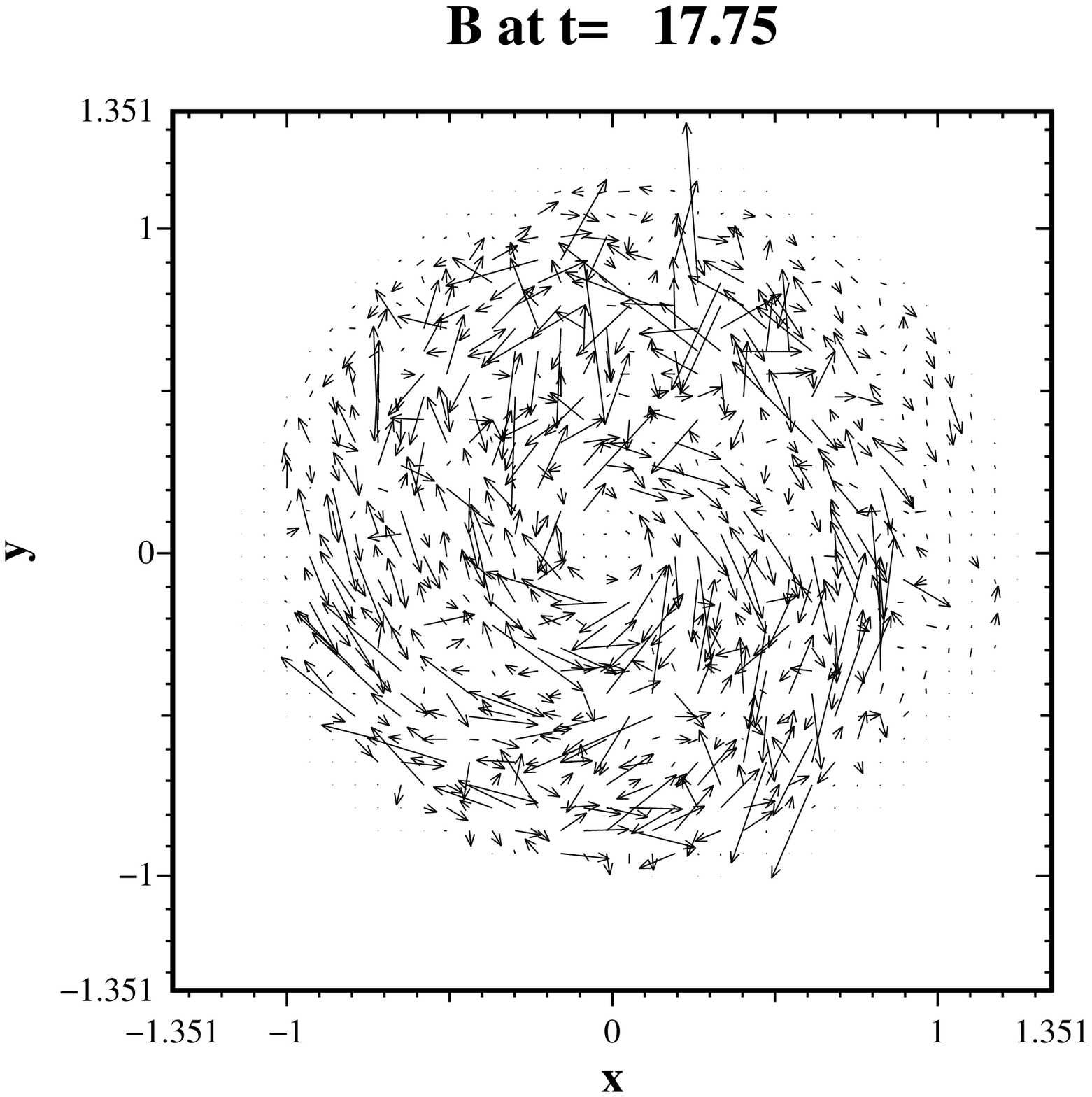}
\caption{Snapshots of field vectors at $t\approx 13.85$ Gyr for (a) Model 203, $q_{\rm i}=2$ and
(b) Model 204, $q_{\rm I}=9$. We note that the scale factor for the vectors in (b)
is half that in (a). }

\label{modr203+204x}
\end{figure}

We also ran cases with $f_{\rm I,2}$ given by Eq.~\ref{enhanc2} and $q_{\rm I}=2, 10$,
with correspondingly larger increases in the global magnetic energy.
The field vectors at $t\approx 13.85$ Gyr with $q_{\rm I}=2, 9$ (Models 203, 204) are shown in
Fig.~\ref{modr203+204x}.

\subsection{Analysis of Results}
\label{analysis}

\begin{figure}
(a)\includegraphics[width=0.80\columnwidth]{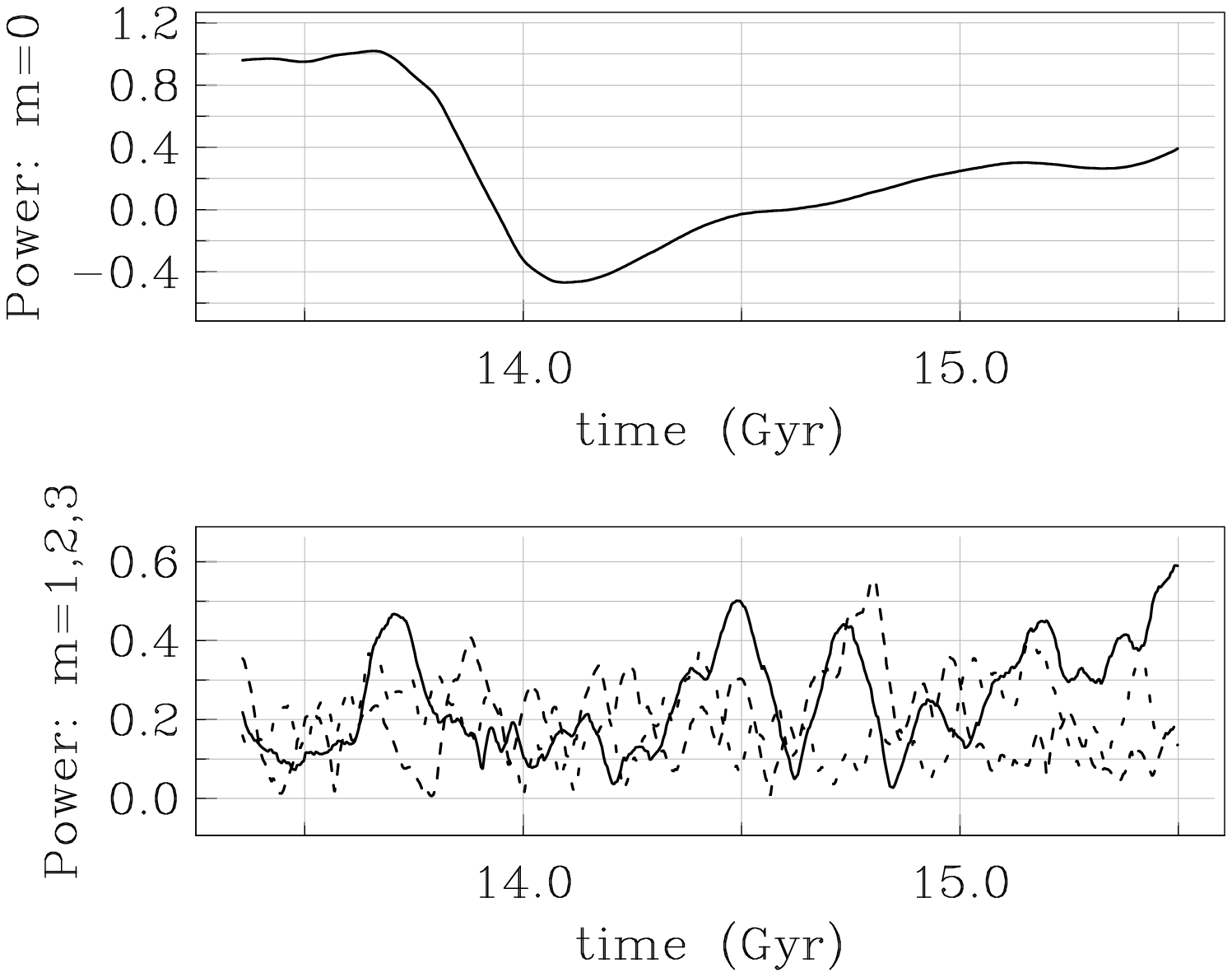}\\
(b)\includegraphics[width=0.80\columnwidth]{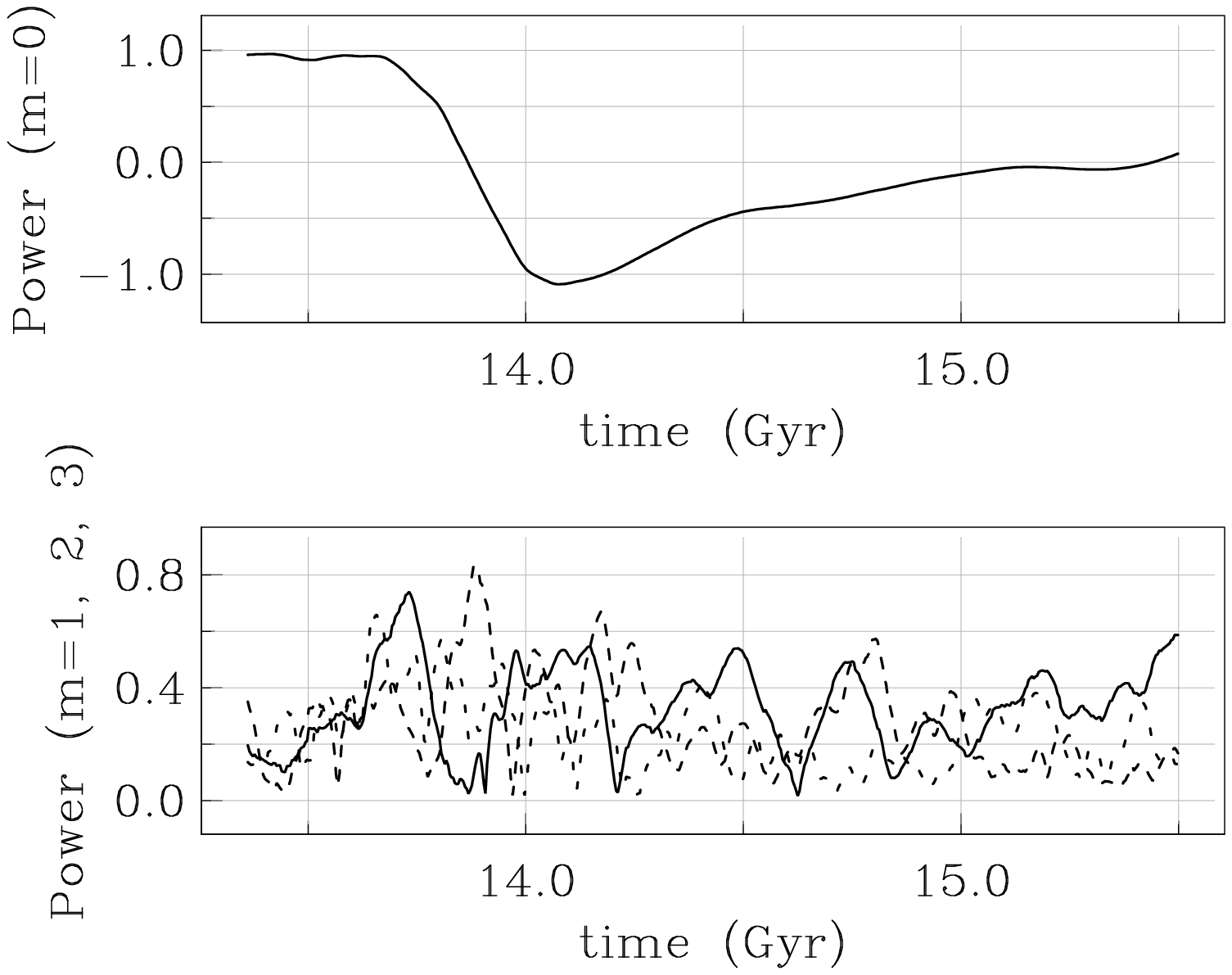}
(c)\includegraphics[width=0.80\columnwidth]{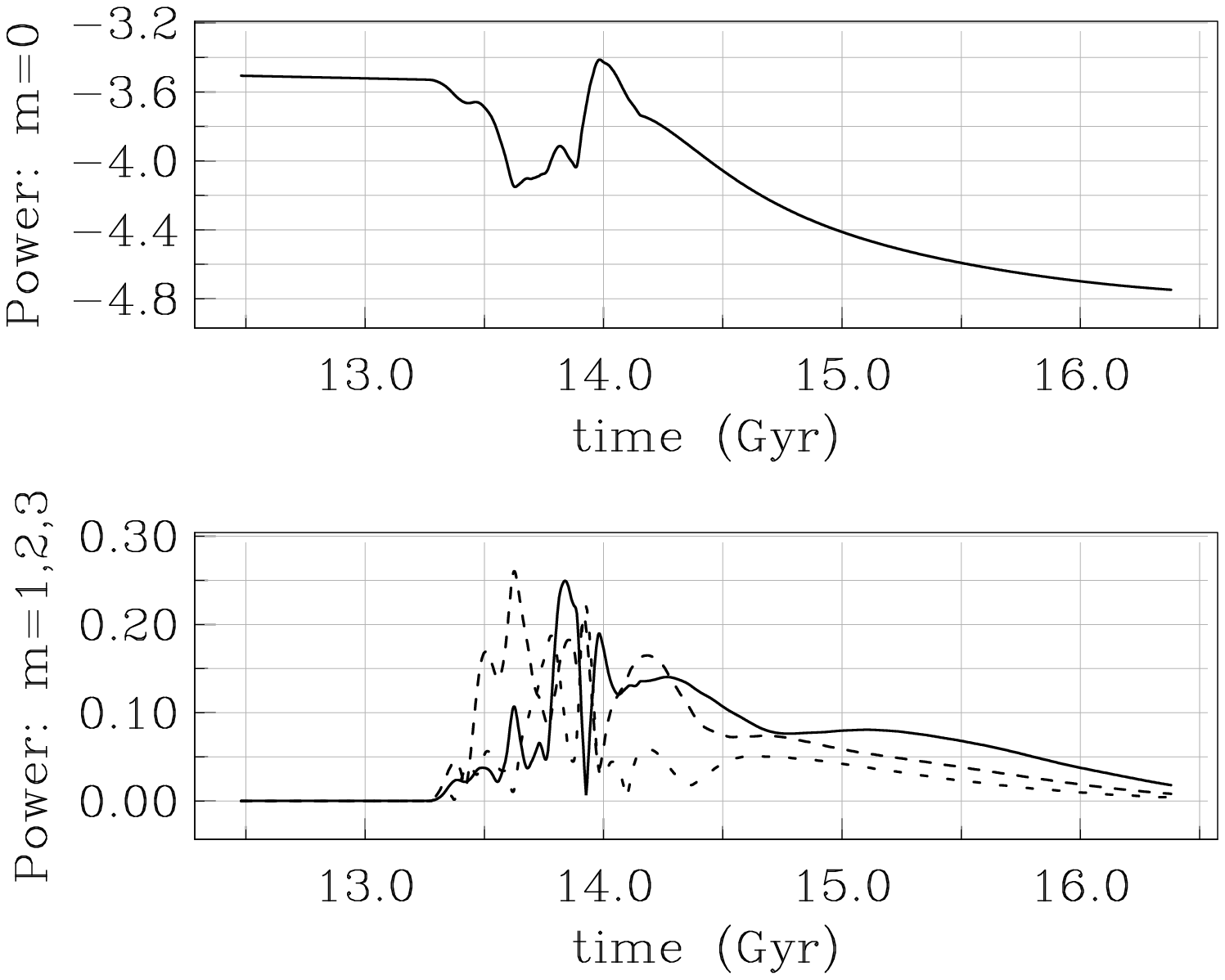}
\caption{Running time averages of
global Fourier integrals $P_i$ of $B_\phi$  through and beyond the interaction,
(which begins
at $t\approx 13.3$ Gyr and ends at $t\approx 14.2 $ Gyr). Upper panels $P_0$, lower panels
$P_1$ (continuous), $P_2$ (broken), $P_3$ (dotted):  (a) Model 16, (b) Model 203, (c) Model 9 (no field injections, data not averaged).
}
\label{fou_sgn_r16x+r203x}
\end{figure}

The effects of the interaction do not show up very dramatically in e.g.
Fig.~\ref{modr16x}, but it is clear from Fig.~\ref{modr16x_tav} that there
are profound effects on the magnetic field. This is in apparent contrast
to the situation shown in Figs.~\ref{modr9x} and \ref{modr9x_t}.
If the fields shown in Fig.~\ref{modr16x} are averaged over a spatial
scale larger than the scale of the injections, the confusing
effects of the small-scale fields are removed, and a significant increase in the azimuthal magnetic field component 
by the non-circular velocities can be seen. However, this field is still turbulent anisotropic (i.e. ordered)\footnote{i.e. appearing ordered in radio polarization observations.},
but the regular large-scale field may be weakened (see Fig.~\ref{fou_sgn_r16x+r203x}a,b) 
as both azimuthal magnetic field directions can be found. These effects
contribute strongly to the increase in global energy through the interaction, 
as seen in
Fig.~\ref{modr16x_tav}. Such an ordered field increases the total as well as the polarized emission. 
However, it can only partly be regarded as being part of the large-scale (regular) 
magnetic field of the disc.

\begin{figure}
(a) \includegraphics[width=0.80\columnwidth]{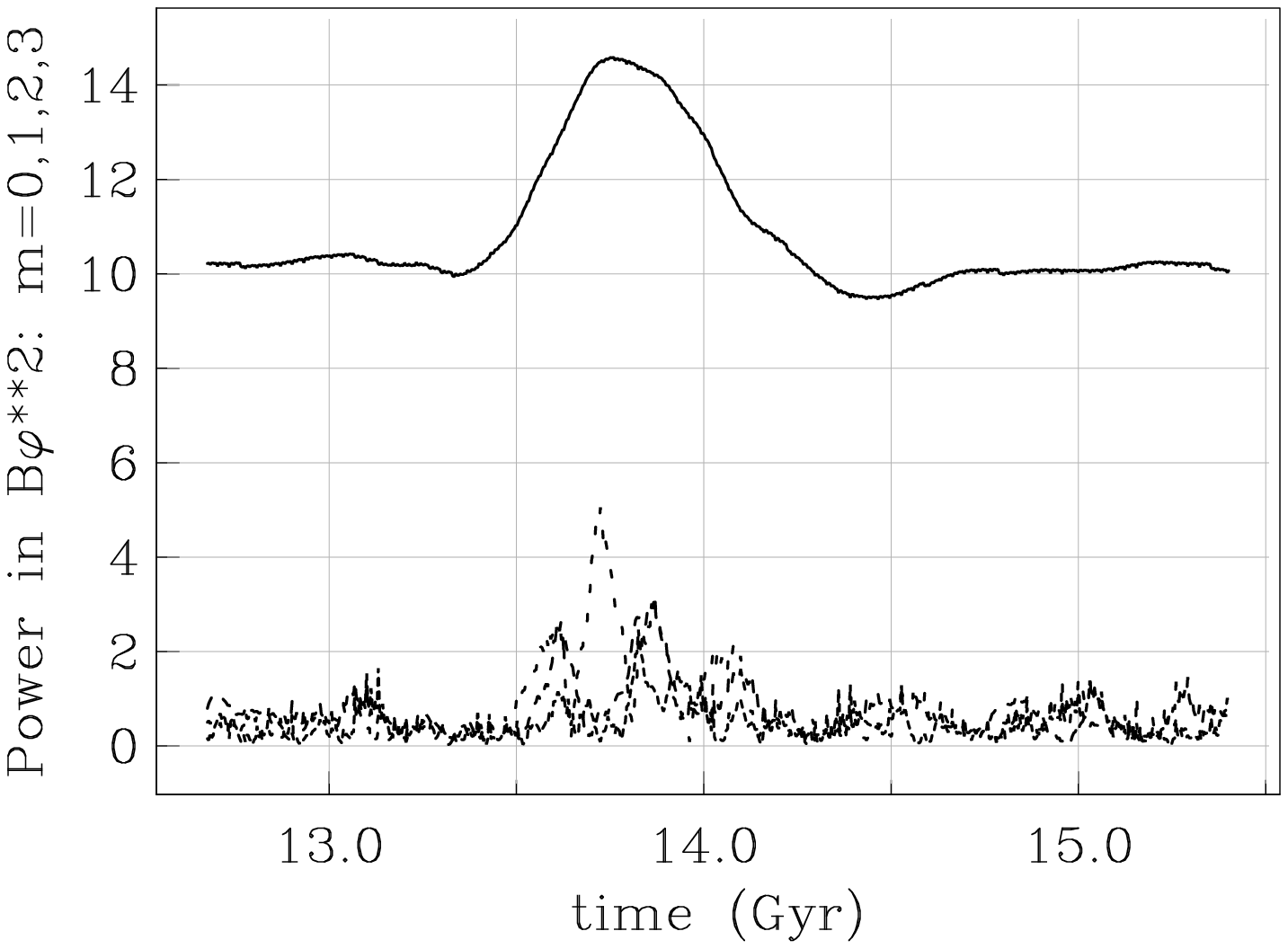}\\
 (b) \includegraphics[width=0.80\columnwidth]{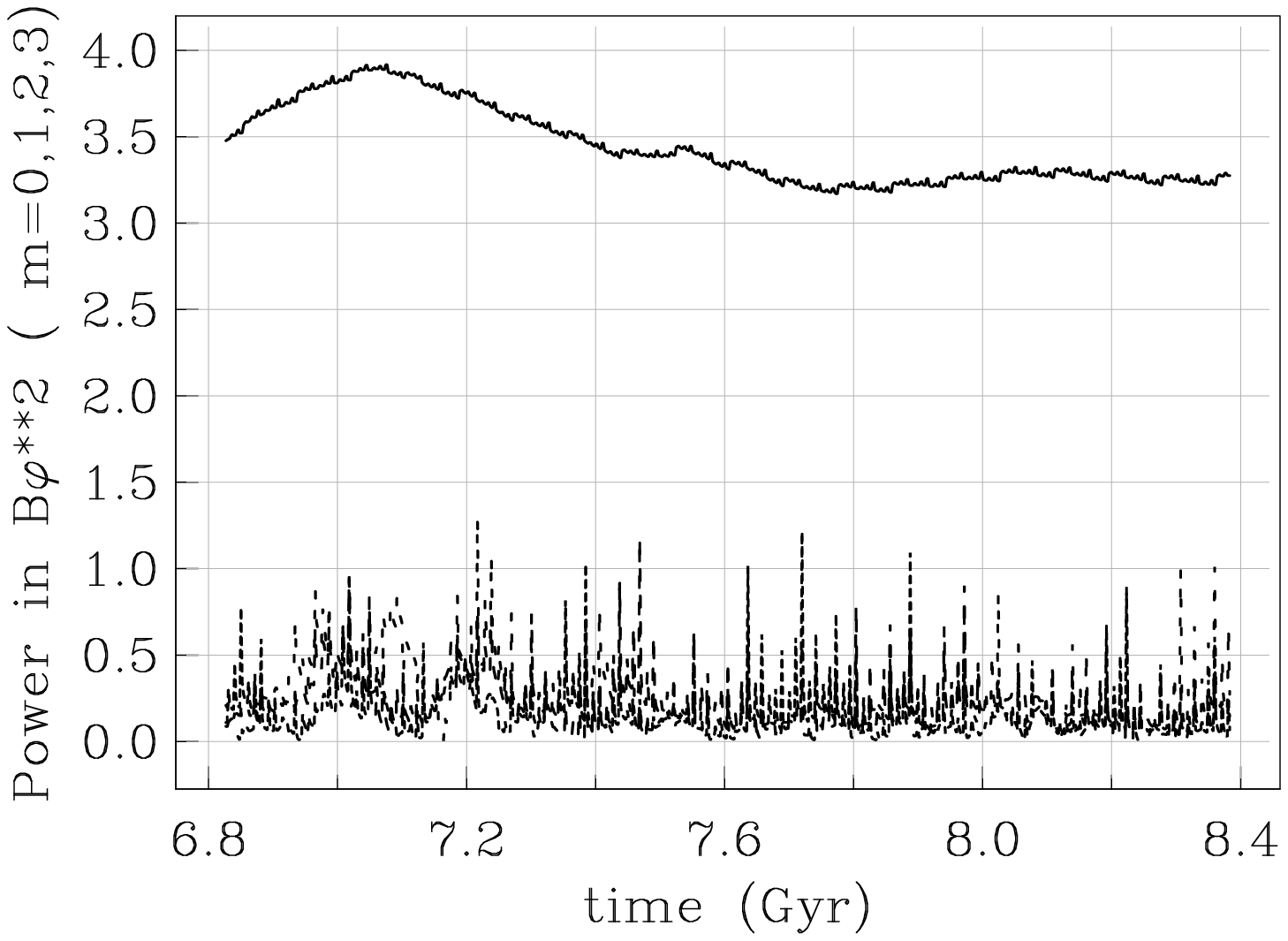}\\
(c)\includegraphics[width=0.80\columnwidth]{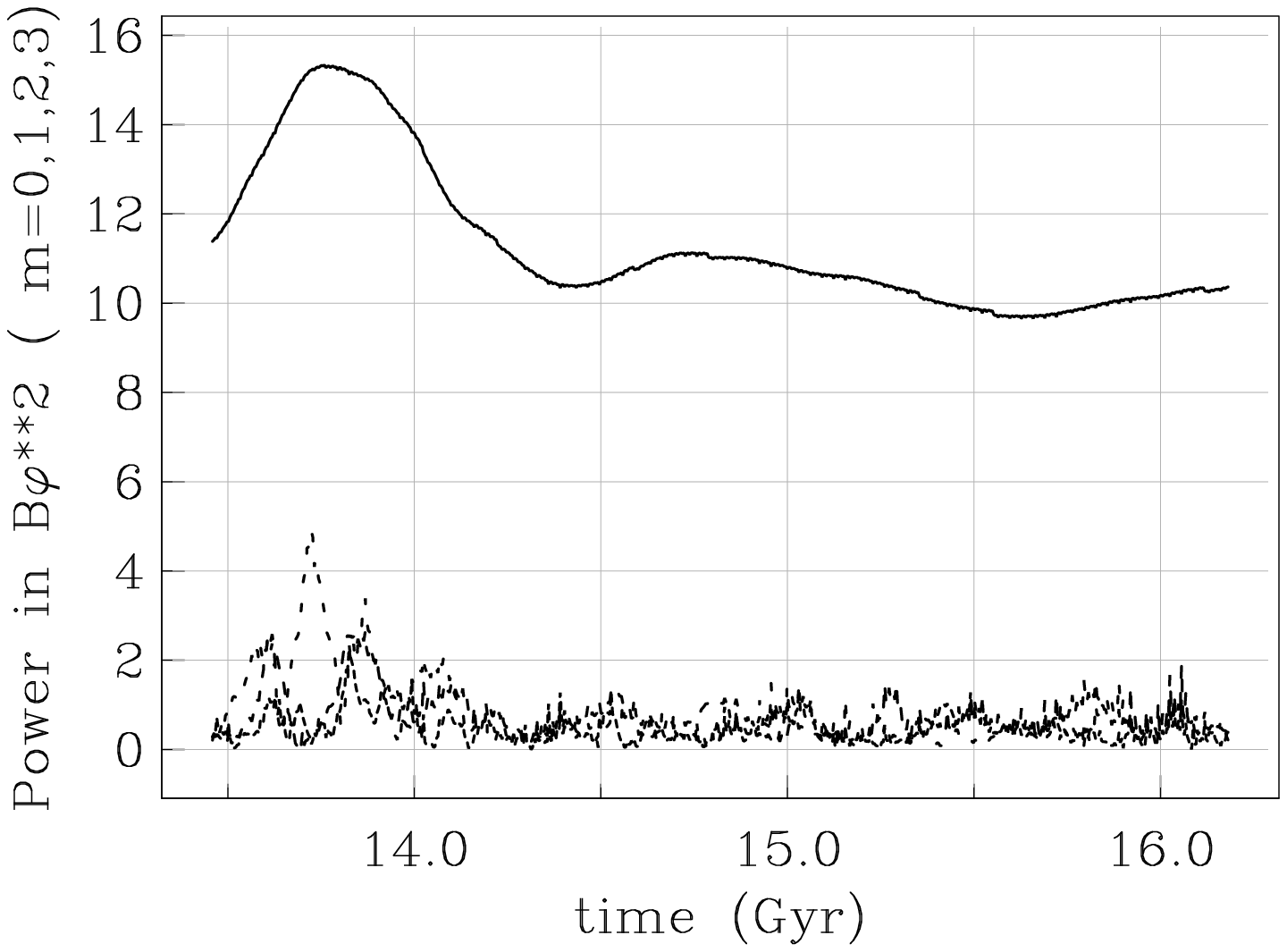}
(d)\includegraphics[width=0.80\columnwidth]{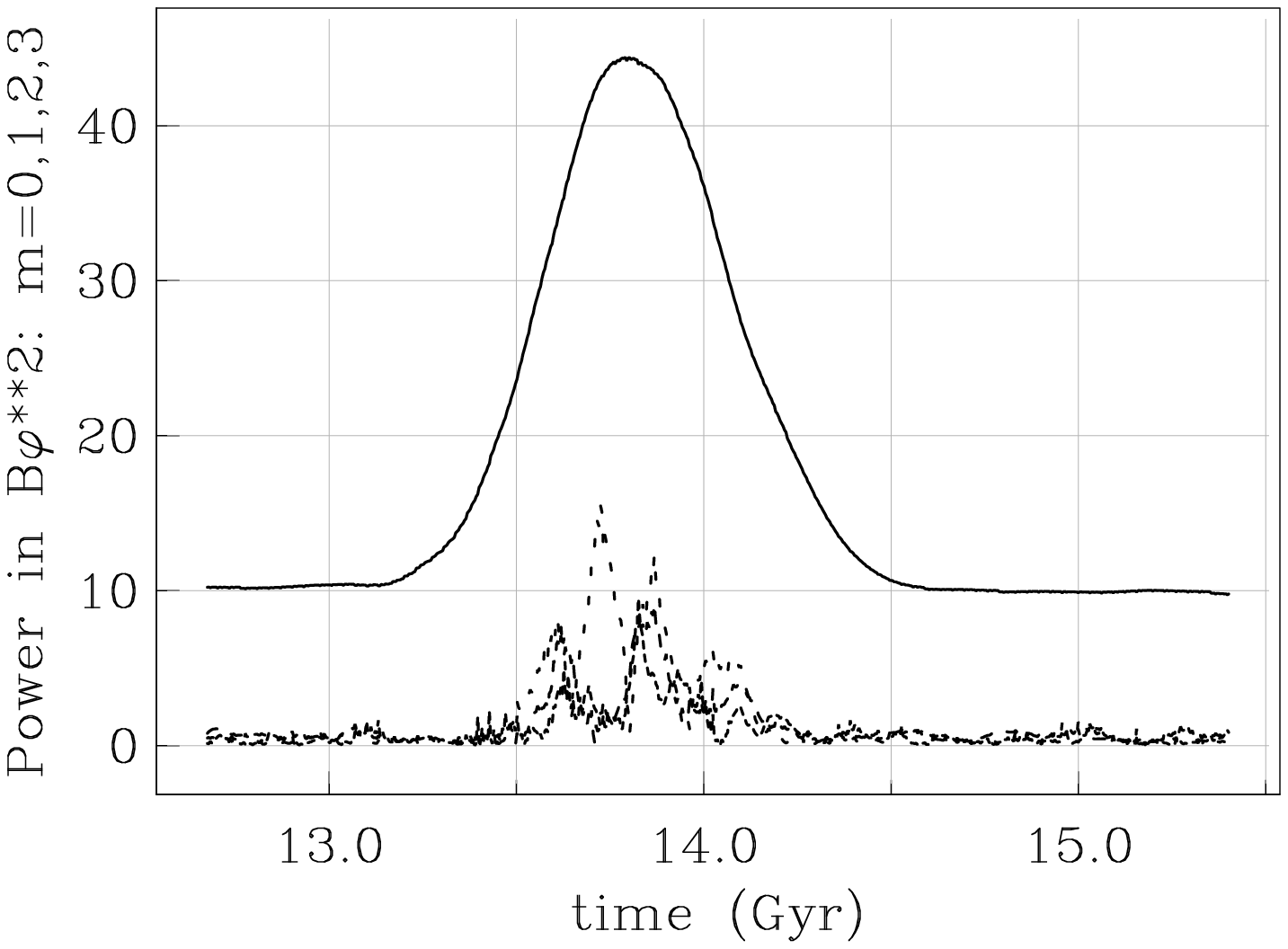}\\
\caption{Time averaged
global Fourier integrals $p_i$ of $B_\phi^2$  through and beyond the interaction
(which begins
at $t=13.3$ Gyr, and ends at $t\approx 14.25$ Gyr in panels (a),
(c), and (d) (Models 16, 101, 201 respectively),
and lasts from approximately from 6.6 Gyr to 7.5 Gyr in panel (b) (Model 17).  In each panel the upper continuous
curves are for $m=0$, and the lower broken, dot-dashed and dotted curves
are for modes $m=1, 2, 3$ respectively.
}
\label{r16x_fou}
\end{figure}

\begin{figure}
(a)\includegraphics[width=0.80\columnwidth]{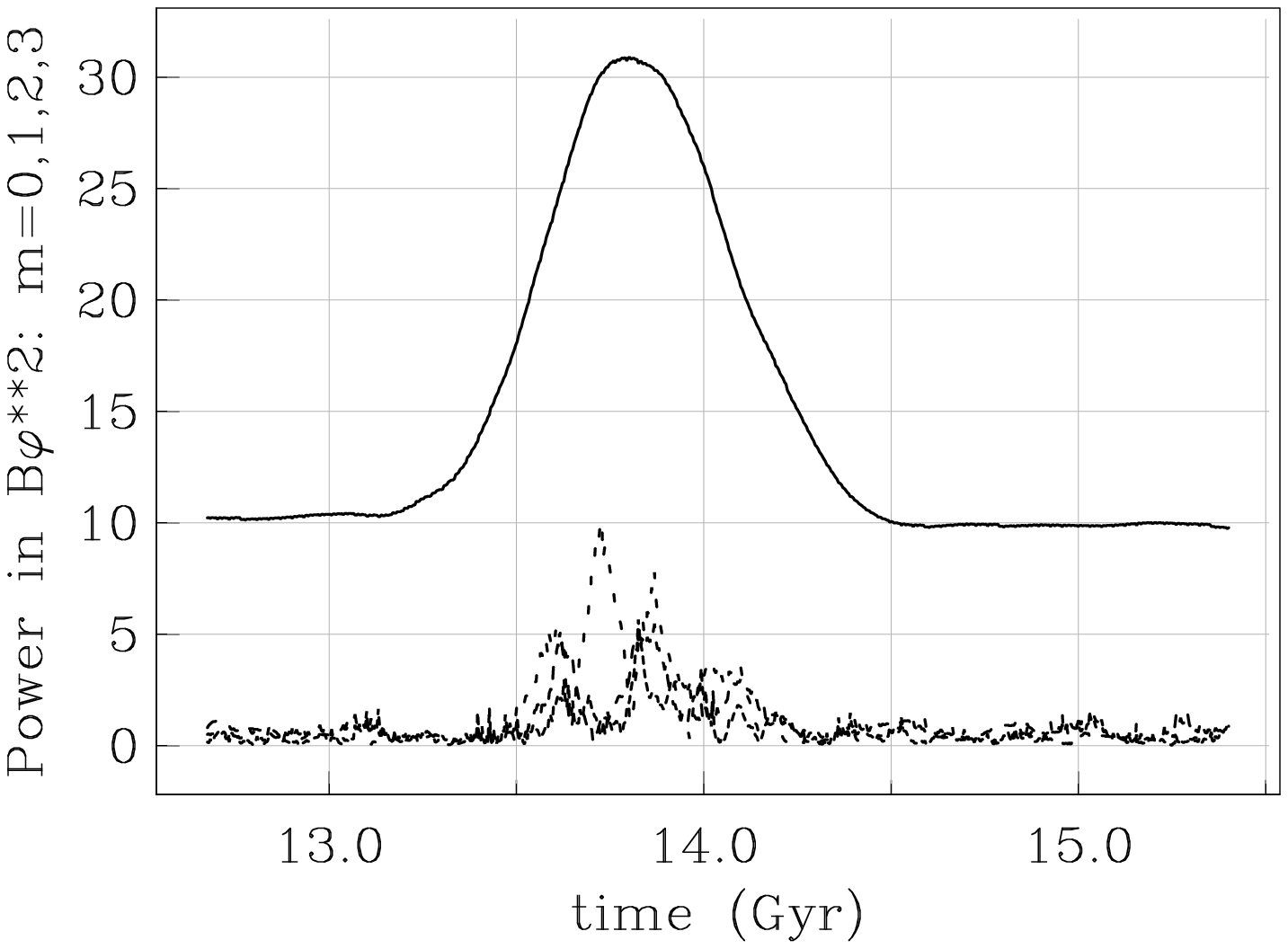}
(b)\includegraphics[width=0.80\columnwidth]{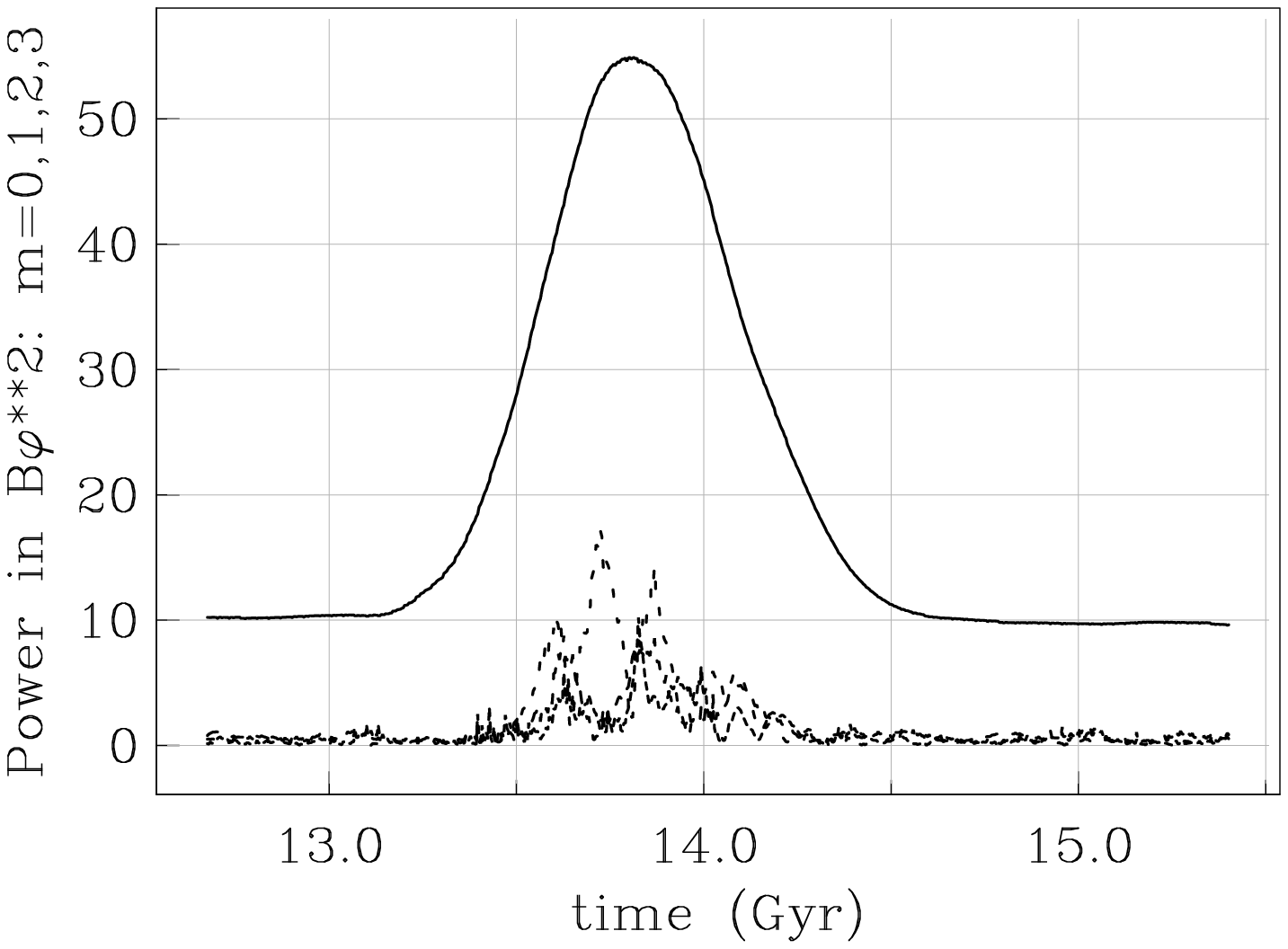}
\caption{Running time averages of
global Fourier integrals $p_i$ of $B_\phi^2$  through and beyond the interaction
(which begins
at $t\approx13.3$ Gyr and ends at $t\approx 14.2 $ Gyr): (a) Model 202 ($q_{\rm I}=1$), (b) Model 203 ($q_{\rm I}=2$).
\protect Notation as in Fig.~\ref{r16x_fou}.
}
\label{enhanced}
\end{figure}

In order to display this more clearly, we computed the global integrals
of magnetic energy in the azimuthal field
$f_m=\int B^2_\phi\cos (m\phi) dxdy$, $g_n=\int B^2_\phi\sin (n\phi) dxdy$, $m=0, 1, 2, 3$,
$n=1, 2, 3$ during the interaction. We then set $p_i=(f_i^2+g_i^2)^{1/2}$, with $g_0=0$.
The results are shown in Fig.~\ref{r16x_fou}
for models 16 and 17 (with $\eta=10^{26}$ and $2\cdot 10^{26}$ cm$^2$s$^{-1}$, respectively), and also for Models 201 and 202.
The physical appearance of the fields (Figs.~\ref{modr16x} and \ref{modr17x})
and the time-averaged energies
(Figs.~\ref{modr16x_tav} and \ref{modr17x_tav}) quite rapidly
approach their pre-interaction states after the interaction has ended.
The effects of the interaction are primarily seen in
the mode $m=0$, as are the gross effects of the field injections, see the upper
(continuous) curves in the panels of
Fig.~\ref{r16x_fou}.
Plausibly this is because the field injections occur on a  much
shorter timescale than that of the interaction,
and also the decay times of the
small-scale field components are relatively short.
Furthermore their spatial scale is smaller than that
of the Fourier modes $m=0-3$ analysed.
We show in Fig.~\ref{enhanced} the comparable plots for Models 202 and 203,
where the increase in energy in mode $m=0$ can be seen when
compared to that of Model 16 shown in Fig.~\ref{r16x_fou}a.

In order to isolate more clearly the effects of the interaction on
the global scale field, we evaluated integrals $F_m, G_n$, and quantities
$P_i$ analogous to the $f_m, g_n, p_i$ defined above, with $B_\phi^2$ replaced by the signed quantity $B_\phi$ in
the integrals.
The integrals $\int B_\phi \cos (m\phi) dxdy$ and $\int B_\phi  (n\phi) dxdy$ are a measure of the regular
(mean) field.
The results for Models 16 and 203 are shown in Fig.~\ref{fou_sgn_r16x+r203x},
together with those for Model 9 (no field injections).
Now there is almost no increase in the magnitude of the quantities $P_0$ during the interaction, and the magnitudes
of $P_1, P_2, P_3$
are comparable. However, there is a long-lived change of sign of $P_0$,
apparently corresponding to the reversals that are discernable in the field vector plots,
and the values of $P_0$ do not quickly revert to their pre-interaction levels 
(i.e. there is a long term effect on the large-scale field).
In effect, the interaction resets the initial state
for the post-interaction
evolution. It is known (e.g. Moss \& Sokoloff 2013), that in a non-linear dynamo
model different initial conditions
can produce different stable states, with or without field reversals,
if the system is initially non-linear.
This is in marked contrast to the situation shown in Figs.~\ref{r16x_fou} and \ref{enhanced}, where $p_0$ quickly
returns to the pre-interaction state. 
It maybe relevant that in Model 9 (with no field injections) there is a long-term
change in the global field structure. The corresponding behaviour of the Fourier modes for this model is
shown in Fig.~\ref{fou_sgn_r16x+r203x}c. The change in power in the mode $m=0$
seems to correspond to the removal of the field reversal in the post 
interaction state. The higher modes are then decaying.

\section{Discussion}
\label{disc}

Both the interaction itself and any increase in SFR associated with the interaction
can increase the overall magnetic energy, due to the effects of 
injected random fields and their subsequent shearing,
that give rise to unpolarized and polarized synchrotron radiation, respectively.
In more detail, our results do not seem particularly sensitive to the chosen
radial distribution of the enhancement of the SFR, as shown by
our experiments with the enhancement function $f(r,t)$ given by Eqs. (\ref{enhanc1}) and (\ref{enhanc2}  with $q_{\rm I}=1$). The maximum enhancement of energy is positively correlated with $q_{\rm I}$.
Observed increases in SFR during encounters are rarely greater than by
a factor of 5 (e.g. Matteo et al. 2008), but it is important to remember
that our parameter $q_{\rm I}$ is only a proxy for the effects of
enhanced SFR on small-scale dynamo action. We note again that any assumed linear relation between $q_{\rm I}$ and SFR is only a crude plausible approximation. It is quite probable that
for high SFR the effects of the individual supernovae explosions can no longer be considered as independent,
and the relation begins to saturate.  We use the parameter $B_{\rm inj0}$
as a proxy for all these possibilities. We cannot address this issue here; clearly
the problem deserves specific investigation (e.g. Geng et al. 2012, A.~Beck et al. 2012). 
We note that our results are broadly consistent with
studies of cosmic ray driven dynamos (e.g. Siejkowski et al. 2014 
and references therein). These studies find that for, modest values of the SFR,
galactic fields are enhanced, but that the dynamo ceases to operate
for very high values of SFR.

If we look at the field vector plots during the interaction, even with large
values of $B_{\rm inj}$ some large-scale order appears visible, generated by the
effects of large-scale velocity shear (especially differential rotation on
the injected field (see e.g. Fig.~\ref{modr203+204x}). In  these models
the magnitude of $<B_\phi^2>$ in the lower modes (here concentrated in $m=0$, 
see Fig.~\ref{r16x_fou}) will be a measure of anisotropic fields and will determine
the level of polarized intensity when observing at a low spatial resolution.

\section{Conclusions}
\label{concl}

Overall we can conclude that an interaction affects the magnetic field configuration.
The effects on the field can take the form of additional reversals of the large-scale magnetic field,
can lead to a concentration of magnetic fields in
rings, and so on. The limiting point, however, is that we only have snapshots of
magnetic field configurations for a few examples, so it is
problematic to isolate effects of interactions in the evolution of galactic magnetic field
configurations; we can play with
the governing galactic dynamo parameters to obtain configurations which more-or-less
resemble fields observed in any particular case.
We note again that we have used an old, low-resolution, dynamical model, 
so our results can only be regarded as generic.

A reasonable way to isolate the effects of interactions observationally
would be by a statistical study of
interacting and non-interacting galaxies.  The immediate feature of interest
in our results is the peak in global
energy in the epoch of encounter. The total magnetic energy increases during an encounter
by 15\%,  80\%,  250\% for Models 17,  101 and 202 respectively, which should lead to a corresponding increase in
the total flux of synchrotron radiation.

\acknowledgements{DS is grateful for  financial support from RFBR under grant 12-02-00170-a and to MPIFR for hospitality. The velocity data was provided by Magnus Thomasson.}

%%%%%%%%%%%%%%%%%%%%%%%%%%%%%%%%%%%%%%%%%%%%%%%%%%%%%%%%%%%%

\end{document}